\documentclass[runningheads]{llncs}
\usepackage[T1]{fontenc}
\usepackage{graphicx}
\usepackage{amssymb}
\usepackage{amsmath}

\usepackage[capitalize]{cleveref}

\newenvironment{judgement}
{\crefalias{equation}{judgement}\begin{equation}}
{\end{equation}}

\crefformat{judgement}{#2\judgementref#1#3}
\Crefformat{judgement}{#2Judgement~(#1)#3}

\usepackage{fancyvrb}

\newenvironment{idrblk}
{ \VerbatimEnvironment
  \begin{small}
  \begin{center}
  \renewcommand{\baselinestretch}{1}
  \begin{BVerbatim}}
{ \end{BVerbatim}    
  \end{center}
  \end{small}}

\newenvironment{codeblksmall}
{ \VerbatimEnvironment
    \begin{footnotesize}
  \begin{center}
  \renewcommand{\baselinestretch}{1}
  \begin{BVerbatim}}
{ \end{BVerbatim}    
  \end{center}
  \end{footnotesize}}

\usepackage{listings}
\usepackage{subcaption}
\usepackage{ebproof}
\usepackage{csquotes}
\usepackage{color}

\newcommand{\qttb}[1]{\ensuremath{\overset{#1}{:}}}

\begin{document}
\title{SynQ: An Embedded DSL for Synchronous System Design with Quantitative Types}
\titlerunning{SynQ}
%
\author{Rui Chen\inst{1}\and
Ingo Sander\inst{1}}
\authorrunning{R. Chen and I. Sander}
%
\institute{School of EECS, KTH Royal Institute of Technology, Stockholm, Sweden\\
\email{\{ruich, ingo\}@kth.se}}
\maketitle              
\begin{abstract}
System design automation aims to manage the design of embedded systems with ever-increasing complexity. 
To the success of system design automation, there is still a lack of systematic and formal design process because an entire design process, from a system's specification to its implementation, has to deal with inherent concerns about the systems' different aspects and, consequently, inherent semantic gaps. 
These gaps make it hard for a design process to be traceable or transparent. 
Particularly, guaranteeing the correctness of produced implementations becomes the main challenge for a system design process.

SynQ (\textbf{Syn}chronous system design with \textbf{Q}uantitative types) is an embedded domain specification language (EDSL) targeting the design of systems obeying the perfect synchrony hypothesis.
SynQ is based on a component-based design framework and, by design, facilitates semantic coherency by leveraging the quantitative type theory (QTT) and language embedding.
SynQ enables a semantically coherent design process, including formal specification and verification, modelling, simulation and code generation.
This paper presents SynQ and its underlying formalism and demonstrates its features and potential for semantically coherent system design through a case study.
\keywords{Synchronous System Design \and Component-Based Design \and Formal Methods \and Domain Specific Language}
\end{abstract}

\section{Introduction}
\label{sec:intro}
The ever-increasing complexity of embedded systems necessitates a more rigorous, systematic and automatic system design process.
The current lack of suitable methods is addressed by recent approaches, such as platform-based design \cite{sangiovanni2001platform}, component-based design, e.g., the BIP (\textbf{B}ehaviour, \textbf{I}nteraction, \textbf{P}riority) framework \cite{gossler2005composition,basu2008bip}, and contract-based design \cite{sangiovanni2012taming}, and investigations on system design activities themselves \cite{sifakis2014toward,sifakis2015system}.
As agreed by this research, embedded system design needs to be pushed to a higher abstraction level and conducted in a top-down manner.
It emerges that bridging the gap between abstraction levels is a critical challenge to be addressed.
Semantic coherency is a particularly challenging problem due to the existence of semantics gaps, which are caused by different concerns on specific aspects at different abstraction levels in the design process.

To address this challenge, we present \textbf{SynQ}, an embedded DSL (EDSL) targeting \textbf{Syn}chronous system design with \textbf{Q}uantitative types.
SynQ is designed and implemented with the consideration of bridging abstraction gaps in a design process by enhancing \textit{semantic coherency}.
Although based on a strong formal foundation, it is, at the same time, practical enough to allow us to generate \textit{executable software} and \textit{synthesisable Verilog HDL code} from synchronous systems modelled in it.

The discussion of this paper is based on a three-stage decomposition of system design processes proposed in \cite{sifakis2014toward}, in which a system design process is decomposed into three critical stages: requirement specification, proceduralisation and materialisation.
SynQ is based on the component-based design framework \cite{sifakis2015system}, which can be seen as a specialisation of the three-stage decomposition.
In the \textit{requirement specification} stage, a system's expected functional behaviour and constraints are specified, which tells \textbf{why} the system is demanded.
Later in the \textit{proceduralisation} stage, the executable model of the system that satisfies specified requirements is built based on a set of atomic (function) components.
This stage explores \textbf{how} a system can be designed.
Finally, in the \textit{materialisation} stage, the model is implemented by implementing function components with physical components.
With extra-functional behaviour introduced by physical components, this is the stage where \textbf{trade-offs} are made to satisfy design constraints.

\subsection{The Need and Challenge of Semantic Coherency}

In pursuit of \textit{productivity}, \textit{correctness} and \textit{automaticity} demanded by system design processes, \textbf{semantic coherency} is recognised as a critical principle and challenge \cite{sifakis2014toward,sifakis2015system}.
Semantic coherency means that a common semantic model shall be shared by languages employed by a design process.
Under the three-stage view, a design process benefits from enforcing semantic coherency in two orthogonal directions.
Horizontally, within each stage, semantic coherency guarantees that the composition of components, regardless of in which language they are specified/implemented, can have a deterministic behaviour which can be reasoned from each component's behaviour and their composition.
This property is a key enabler for \textit{component-based construction} and \textit{correctness-by-construction} because it ensures that unexpected behaviour will not be introduced when constructing systems by composing sub-systems.
Vertically, a unified semantic model bridges gaps between abstraction levels entailed by different stages.
That is to say, with coherent semantics, artefacts at different stages can be formally/rigorously related by their semantics, facilitating \textit{traceability} and \textit{correctness-by-construction} desired by design processes \cite{sifakis2015system}.

\begin{figure*}
    \centering
    \includegraphics[width=\linewidth]{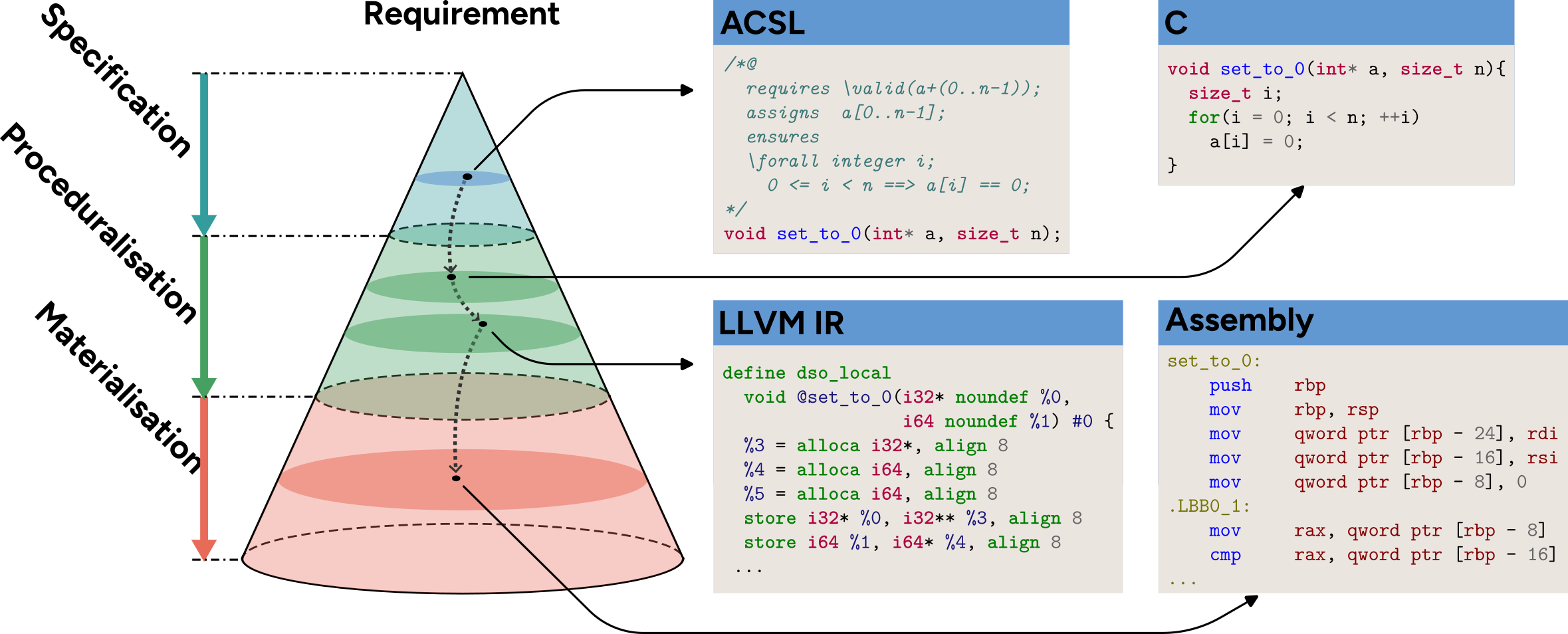}
    \caption{The visualisation of a top-down design flow under the three-stage decomposition view in which each plane corresponds to a design space at a specific abstraction level and exemplified artefacts that can will be produced in a C-centric design flow.}
    \label{fig:design-flow-c}
\end{figure*}

On the other hand, achieving semantic coherency in a system design process remains challenging.
Because a system-level design process involves considerations of multiple aspects at different abstraction levels.
Each of these aspects and abstraction levels forms its own problem domain and desires a dedicated domain-specific language (DSL) with carefully selected semantics.
Only in this way can solutions be made with maximised productivity and correctness guarantee.
For instance, considering a C-centric design process illustrated in \Cref{fig:design-flow-c} using C as the behaviour modelling language, neither specifications nor models that are ready to be optimised or implemented are suitable to be given as C programs.
Instead, the ANSI/ISO C Specification Language (ACSL) \cite{baudin2021acsl} can be employed as an appropriate specification language. 
Meanwhile, it is more practical that optimisations and implementations are conducted based on an intermediate representation (IR), e.g. the LLVM IR \cite{lattner2004llvm}.
With these languages involved, a C-based behaviour model is, in essence, associated with two semantics: \textit{axiomatic semantics} \cite{hoare1969axiomatic} employed for verifying the model with respect to its ACSL specification and \textit{operational semantics} \cite{plotkin1981structural} entailed by translating this model into the selected IR.
Even though there exists some effort to relate these two semantics, e.g. \cite{schafer2016axiomatic}, none of these semantics can easily replace each other because they focus on two distinct aspects, logic and computation, respectively.

To improve semantic coherency, one possible way is to introduce a general-purpose language as the meta-language and interpret DSLs involved in a design process by the meta-language.
The interpretation can be conducted by either a standalone program or a program in the meta-language.
The former is often found in modern compilation frameworks.
In such a case, the standalone program is often referred to as a language frontend, and the meta-language is the internally used IR.
This approach has indicated its potential for hosting ubiquitous source languages through many successful projects, e.g., the Java Virtual Machine (JVM) \cite{lindholm2013jvm} and the LLVM compilation framework \cite{lattner2004llvm}.
Meanwhile, it can also target heterogeneous platforms and introduce multiple abstraction levels (c.f. the MLIR project \cite{lattner2021mlir}).
The latter is recognised as \textit{language embedding} \cite{gibbons2014folding}, in which a DSL is specified and implemented/interpreted by a set of definitions in the meta-language.
The language embedding approach has been employed to formalise languages in theorem prover, e.g., the HOL-ELLA, HOL-SILAGE and HOL-VHDL project(s) \cite{boulton1992experience} that embeds hardware description languages (HDLs) into HOL \cite{gordon1993hol} and the Vélus project \cite{bourke2017formally,bourke2023verified} that formalises and bridges Lustre \cite{pilaud1987lustre} and Obc \cite{biernacki2008obc} in the Coq theorem prover \cite{coq1996coq}.
In the meantime, language embedding can also host DSLs in practice. 
For instance, several real-life languages, e.g., Lustre, Simulink and nesC, have been embedded in the BIP framework \cite{bozga2009modeling,sfyrla2010compositional,basu2007using}.

However, when it comes to system design, semantic coherency introduced by these practices cannot cover all stages mentioned earlier.
The closest one is the Vélus project that related declarative, dataflow-based semantics \cite{pilaud1987lustre} with operation semantics implied by a sequence of imperative instructions.
The Vélus project, however, is still limited in the proceduralisation stage, which cannot directly bridge the semantic gap between the specification and the proceduralisation stage illustrated in the early example.

\subsection{The SynQ Language}

As introduced earlier, \textbf{SynQ} is an embedded DSL (EDSL) targeting Synchronous system design with Quantitative types, which is designed to facilitate a semantically coherent system design process in a component-based design framework \cite{sifakis2015system}.
By synchronous systems, we mean here the same subset of systems, mainly embedded control and signal processing systems, that can be modelled by a synchronous programming language, e.g., \cite{pilaud1987lustre,benveniste1991signal}.
Such systems are often safety-critical and reactive, desiring deterministic execution time and memory-bound.
These requirements necessitate a more rigorous, traceable and systematic design process, which motivates this paper.

Semantic coherency is assured by the design and implementation of SynQ as follows: 
\begin{itemize}
    \item the carefully selected meta-language Idris2 \cite{brady2021idris} in which specification, verification, and computation can be seamlessly conducted; and 
    \item a systematic embedding of the DSL by the tagless final approach \cite{carette2009finally}, which allows us to fully leverage the meta-language and employs a unified way to model and specify aspects of artefacts and design steps involved in a design process.
\end{itemize}
Idris2 \cite{brady2021idris} is a Haskell-like, general-purpose functional language typed by the quantitative type theory (QTT) \cite{mcbride2016got,atkey2018syntax} that combines dependent and linear types.
Functional languages are recognised for their minimality, inherent parallelism (the Church-Rosser theorem) and tight relation with formal methods (the Curry-Howard correspondence) \cite{nederpelt2014type}.
Among these advantages, the Curry-Howard correspondence is a key enabler for coherency.
By the Curry-Howard correspondence, theorems (specifications) and their proofs directly correspond to types and programs, respectively.
In this way, the logic aspect, i.e. specification and reasoning, is related to the computation aspect (modelling) concerned in a design flow within the same language, just as we expected in the early example.
Specifically, QTT enables us to specify fine-grained properties of systems in types, which can be either functional or extra-functional.
For instance, in QTT, we can soundly and completely type programs with polynomial time computation, which is, in essence, an extra-functional property of programs \cite{atkey2024polynomial}.
Further, recent investigations have also indicated that the application of QTT is not limited to software developments.
It can also be used to type HDLs \cite{de2023wiring} or semi-formally host a component-based design framework \cite{chen2024qtt}.
These facts make Idris2 an ideal meta-language for system design processes.

The \textit{tagless final approach} \cite{carette2009finally} is proposed to embed \textit{typed} object-language without encountering the tag problem (c.f. Section 1.1 in \cite{carette2009finally}).
In contrast to a conventional deep (initial) embedding that specifies the object-language's syntax as a type in the meta-language, the tagless final approach starts by specifying constructors consisting in the object-language as a set of type signatures of polymorphic functions.
The set of signatures is often referred to as \textit{symantics} in literature as it constrains both \textbf{sy}ntax and se\textbf{mantics} of the object-language.
Any \textit{consistent} implementation of such a set of type signatures forms an interpretation, and hence, an interpreter, of the object-language.
The shallow nature of the signature-based specification enables us to fully utilise the infrastructure of the meta-language, especially its type system.
Further, an interpretation of an object-language can be made in either a context-free or context-sensitive manner or to target another object-language.
This flexibility allows models of an artefact's different aspects and abstraction levels that occur in a design process to be captured and specified in a unified form in the meta-language.
Thus, semantic coherency is facilitated.

\subsection{Contributions and Contents}

We summarise the contributions of this paper as follows:
\begin{itemize}
    \item this paper presents SynQ, an EDSL for synchronous system design in the component-based construction framework that leverages quantitative types for system constructions (\Cref{sec:sym,sec:interpreter});
    
    \item this paper showcases that a wider range \textit{semantic coherency} can be achieved by systematically embedding DSLs, through the tagless final approach, into the meta-language Idris2 that implements an advanced type system (QTT) (\Cref{sec:interpreter,sec:examples});

    \item this paper demonstrates with practical examples that SynQ enables a design process that can target both software and hardware by simply interpreting terms differently and enables us to conduct specification, formal verification by theorem proving, and design transformations \textit{seamlessly} (\Cref{sec:interpreter,sec:examples});
    
    \item the implementation of SynQ with all examples used in this paper publicly available at \cite{omitted}.
\end{itemize}
Before the contributions above are presented, background knowledge, e.g., the QTT, programming in Idris2 and the tagless final approach, will be briefly reviewed in \Cref{sec:preliminary}.
Related works that are not covered in the introduction and further discussions will be presented in \Cref{sec:related} and \Cref{sec:conclusion}, respectively.

\section{Preliminary}
\label{sec:preliminary}
In this section, we briefly review the \textit{quantitative type theory}, including dependent and linear types, and the \textit{tagless final approach}, which are employed as the basis of this paper.
The presentation will mainly be made based on code in Idris2 with a special focus on their \textit{practical aspect}, which is aligned with the introduction of SynQ in later sections.

\subsection{Quantitative Type Theory}

Quantitative type theory is initially proposed by McBride \cite{mcbride2016got} to harmoniously combine \textit{dependent types} \cite{nederpelt2014type} and \textit{linear types} \cite{girard1987linear,abramsky1993computational}.
The combination allows us to benefit from both the ability of \textit{specification} and \textit{verification} enabled by dependent types and \textit{fine-grained resource control} enabled by linear types.
To this end, we can specify and verify \textit{intrinsic properties} of programs in QTT as it has been shown in \cite{atkey2024polynomial}.

\subsubsection{Dependent Types}

In type theory, the fundamental problem to be answered is \textit{whether a term \(x\) can be typed with a type \(T\) in a context where a sequence of terms \( x_1, x_2, \dots, x_n\) are respectively typed with types \( T_1, T_2, \dots, T_n\)}.
This problem is recognised as a \textit{judgement} and often stated in the form:
\[ x_1: T_1, x_2 : T_2, \dots, x_n: T_n \vdash x:T, \]
in which atoms of the shape $\alpha : \tau $ are \textit{type bindings} that bind the type $\tau$ to the term $\alpha$. 
Terms on the LHS of $\vdash$, which forms the \textit{context}, could be used to construct the RHS of $\vdash$ and the rest of the context.

According to the Curry-Howard correspondence, a judgment can be interpreted from both the logical and the computational side.
From the \textit{logic} perspective, a type \(T\) forms a \textit{theorem}, and a term bound with it encodes a \textit{proof} of the theorem.
A judgment then states that \textit{$x$ is a proof of the theorem $T$ in the context where theorems \( T_1, T_2, \dots, T_n\) are all proved}.
From the \textit{computation} perspective, types and terms are interpreted as types in a programming language and programs written in the language, respectively.
Hence, a judgment is interpreted as that \textit{a program $x$ is of the type $T$ if sub-programs \( x_1, x_2, \dots, x_n\) consists in $x$ are of types \( T_1, T_2, \dots, T_n\), respectively.}\
SynQ leverages this possibility of interpreting typed terms differently to bridge the gap between the specification and proceduralisation stages.

A conventional programming language has two disjoint sets of symbols to represent terms and types.
In such a case, terms cannot be referred to in types, and hence, the properties of terms cannot be specified and verified.
Dependent types address this problem by mixing terms and types, i.e., allowing types \textit{dependent} on terms and vice versa.
For instance, dependent types allow us to specify the following judgement:
\begin{judgement}
\label{judg:vec-len}
    a : \mathtt{Type}, n : \mathtt{Nat} \vdash \text{\textit{len}} : \mathtt{Vect}\: n \: a \to \mathtt{Nat},
\end{judgement}
in which $n$ is a term of the type $\mathtt{Nat}$ (natural numbers) and \(\mathtt{Vect}\: n \: a\) is a type (vectors of length $n$ w/ elements of the type $a$) dependent on the value of $n$, e.g., \(\mathtt{Vect}\: 3 \: a\) is the type of all vectors of length three, which \textit{cannot} be used to type a vector with two elements only.

From the computation perspective, \Cref{judg:vec-len} \textit{claims} that we can construct a function named \textit{len} from vectors of length $n$ to natural numbers in the context where $n$ is a natural number and $a$ is a type.
It corresponds to a function declaration in Idris2:
\begin{idrblk}
len: Vect n a -> Nat
len = ?len_impl
\end{idrblk}
in which \verb|?len_impl| states a \textit{hole} where the implementation have not be given.
Idris2 allows us to inspect the context and the target type of a hole.
An inspection of \verb|?len_impl| will give us:
\begin{idrblk}
`--             0  a : Type
                0  n : Nat
     -----------------------------------
      "Main.len_impl" : Vect n a -> Nat
\end{idrblk}
which directly matches \Cref{judg:vec-len}.
The 0 before $a$ and $n$ denotes how many times the term should be used, which will be explained later.

With dependent types, we can also introduce a type dependent on \textit{len} to specify its expected behaviour and verify if an implementation satisfies the specification.
For instance, let \textit{len} be the function that returns the length of the given vector.
Its behaviour can be specified by the type in Idris2 that\footnote{Here we omitted the judgement-style specification in a similar form as \Cref{judg:vec-len} because it will require a systematic definition of bindings. The interested reader is referred to Chapter 5 of \cite{nederpelt2014type} for a detailed introduction.}: 
\begin{idrblk}
len_prop: (n: Nat) -> (xs: Vect n a) -> (len xs) = n
\end{idrblk}
which specifies that for arbitrary natural number \verb|n| and vector \verb|xs| of length $n$, the result of applying \verb|len| on \verb|xs| is equal to \verb|n|.
Note that \verb|(len xs) = n| is a dependent type constructed by the type constructor \verb|(=)|, which enables us to assert equality by \textit{reflexivity}.
An implementation of \verb|len| based on what \verb|len_prop| can be implemented (proved), and the corresponding proof of \verb|len_prop| are shown below:
\begin{idrblk}
len: Vect n a -> Nat
len = \x => case x of 
              []        => 0
              (x :: xs) => 1 + len xs

len_prop: (n: _) -> (xs: Vect n a) -> (len xs) = n
len_prop 0 [] = Refl
len_prop (S k) (x :: xs) = cong (1 +) (len_prop k xs)
\end{idrblk}
in which both \verb|len| and \verb|len_prop| are implemented by inductively matching the structure of the given vector.
The second line of \verb|len_prop| checks that \verb|Refl| (reflexivity) can be typed with \verb|len [] = 0| because its LHS can be reduced to 0 (by the first case entry of \verb|len|) and \verb|Refl: 0 = 0| is well-typed.
The last line of \verb|len_prop| aims to prove \verb|len (x :: xs) = (S k)| in which \verb|S| is the data constructor of \verb|Nat| meaning the successor of \verb|k|, which basically means the same of \verb|1 + k|.
By reducing \verb|len (x :: xs)| to \verb|1 + len xs|, the target term to be constructed is of the type \verb|1 + len xs = 1 + k|.
In \verb|len_prop|, the term is constructed based on the fact that equality is a congruence relation, reflected by \verb|cong|.
That is, \verb|1 + len xs = 1 + k| can be constructed if \verb|len xs = k| can be constructed.
Since terms on both sides in the latter case are always smaller than terms in the former case, we can safely apply \verb|cong| until \verb|0 = 0| is reached.
It then proves that a term of the type \verb|1 + len xs = 1 + k| can always be constructed, and hence, \verb|len (x :: xs) = (S k)| is proven.

It is clear that the implementation of \verb|len| and \verb|len_prop| is \textit{not unique}.
For instance, the dependent type system has the following derivation rule:
\begin{center}
\begin{prooftree}
     \hypo{\Gamma, a : A \vdash b: B}
    \infer{1}[($\to\text{Intro}$)]{\Gamma \vdash \lambda a. b : A \to B}
\end{prooftree}    
\end{center}
which indicates that if $b: B$ hold in a context \((\Gamma, a : A)\), in which $\Gamma$ is an arbitrary sub-context, then we can get a term of type $A \to B$ in the context \(\Gamma\) by abstracting $a$ from $b$.
By applying this rule to \Cref{judg:vec-len}, we can get another \textit{len} function (the former one is renamed to \textit{len'}):
\begin{center}
\begin{prooftree}
     \hypo{ a : \mathtt{Type}, n : \mathtt{Nat} \vdash \text{\textit{len'}} : \mathtt{Vect}\: n \: a \to \mathtt{Nat}}
    \infer{1}[($\to\text{Intro}$)]{ a : \mathtt{Type} \vdash \text{\textit{len}} : (n : \mathtt{Nat}) \to  \mathtt{Vect}\: n \: a \to \mathtt{Nat},}
\end{prooftree}    
\end{center}
In Idris2, the corresponding function declaration and its trivial implementation could be:
\begin{idrblk}
len: {n: Nat} -> Vect n a -> Nat
len {n} xs = n
\end{idrblk}
in which \verb|{n: Nat}| indicates that \verb|n| is an \textit{implicit} argument that can be inferred from the context because whenever a vector is given, so does its length.
With the implementation above, \verb|len_prop| can be simply proved by/implemented as:
\begin{idrblk}
len_prop: (n: _) -> (xs: Vect n a) -> (len xs) = n
len_prop n xs = Refl
\end{idrblk}
From a design process point of view, this non-uniqueness forms the \textit{design space} determined by a type-specified property.

\subsubsection{Linear and Quantitative Types}
\label{sec:preliminary:qtt}
Dependent types enable us to specify \textit{observable behaviours} of functions in types.
That is, a specification of a function $f$'s property is, in general, of the form that if an input $x$ satisfies a predicate $P$, then $f \: x$ satisfies the predicate $Q$ (formally, \(P(x) \implies Q(f \: x)\)).
However, a system design process should also be able to take the \textit{intrinsic properties} of functions/components into account.
For instance, when a platform with limited resources is given, designers should, to some extent, be able to answer the question of whether a given function can be implemented on the platform with these resources.

\textit{Linear types} \cite{girard1987linear,abramsky1993computational} is a type theory that allows us to specify and reason resources-related properties.
Compared to a conventional type theory, following structural rules for deriving types in linear type theory do \textit{not} exist:
\begin{center}
\begin{minipage}{.6\textwidth}
\begin{center}
\begin{prooftree}
     \hypo{\Gamma, x : A, y : A \vdash t: C}
    \infer{1}[(Contraction)]{\Gamma, z:A \vdash t[z/x, z/y]: C}
\end{prooftree}    
\end{center}
\end{minipage}%
\begin{minipage}{.4\textwidth}
\begin{center}
\begin{prooftree}
     \hypo{\Gamma \vdash t: B}
    \infer{1}[(Weakening)]{\Gamma, z: A \vdash t: B}
\end{prooftree}    
\end{center}
\end{minipage}%
\end{center}
The \textit{contraction} rule enables different sub-terms ($x$, $y$) of the same type in $t$ to be substituted by a term $z$ of the same type.
Meanwhile, the \textit{weakening} rule allows an \textit{unused} term to occur in constructing another term.
By prohibiting these rules, a term in linear type theory can be used exactly once, i.e., a term can neither be duplicated nor eliminated (consumed) due to the lack of contraction and weakening rule, respectively.
To this end, resource occupation can be easily specified in linear type theory.
It is worth mentioning that a type theory in which some structural rules are missing is recognised as a \textit{substructural type theory}.
Some of such type theories have already been used in a more practical environment, e.g. the \textit{affine types} employed by Rust \cite{tov2011affine-types}.

The conventional combination of linear and dependent types enforces linearity in \textit{term} constructions and allows terms to be used unrestrictedly in types.
\textit{Quantitative type theory} \cite{mcbride2016got,atkey2018syntax} achieves this combination by annotating each term with notations of how they can be used \cite{petricek2014coeffects,ghica2014bounded,terui2001light}.
In QTT, the usage information of terms is denoted on type bindings.
That is, a judgement in QTT is of the form:
\[ x_1 \qttb{\rho_1} T_1, x_2 \qttb{\rho_2} T_2, \dots, x_n \qttb{\rho_n} T_n \vdash x \qttb{\sigma} T, \]
in which \(\rho_k\) (for \(k \in [1, n]\)) and \(\sigma\) denote how a term should be used.
To be used for usage accounting, $\rho$ and $\sigma$ are elements of the carrier set \(R\) of a \textit{semiring} \((R, +, \cdot, 0, 1)\) in which $+$ is used to sum up the usage of terms and $\cdot$ is used to count nested usage of terms, e.g., terms used in nested function calls.

Specifically, Idris2 employes the \(\{ 0, 1, \omega\}\) semiring (\(R = \{ 0, 1, \omega\}\)), where for all $\rho \in R$, we have \(\rho + \omega = \omega\) and \(\omega\cdot \omega = \omega\).
With this semiring, a term in Idris2 can be used zero times (only in types), precisely once (linearly) or arbitrarily many times (unrestricted).
These usage annotations are named \textit{multiplicities}.
The following presents the type signature with explicit multiplicities annotations of \textit{another} possible variant of the \verb|len| function
\begin{idrblk}
len: (0 a: Type) -> (1 n: Nat) -> (xs: Vect n a) -> Nat
len a n xs = ?len_rhs
\end{idrblk}
in which 0 and 1 denote terms that are to be used only in types and exactly once, respectively, and \verb|xs| has multiplicity $\omega$ that need not be denoted.
As discussed earlier, the usage of \verb|a| and \verb|n| in the type \verb|Vect n a| is not restricted by their multiplicities. 
The inspection of the context of \verb|?len_rhs| is presented as following
\begin{idrblk}
`--              1  n : Nat
                 0  a : Type
                   xs : Vect n a
    -----------------------------
       "Main.len_rhs" : Nat
\end{idrblk}
which indicates that \verb|n| should be used once at the hole and \verb|a| should not be used, while \verb|xs| can be used unrestrictedly.

A less obvious fact is that \textit{all} previous implementations of \verb|len| are valid, even though \verb|n| has multiplicity 1 and is not explicitly used in the first implementation.
It is because of that a linear term can be consumed by \textit{pattern matching}.
Meanwhile, since the type of \verb|xs| depends on the value of \verb|n|, pattern matching on \verb|xs|, which discards \verb|xs| and introduces a value or sub-term(s) of it, will also cause a pattern matching on \verb|n|, as shown by the following:
\begin{idrblk}
len: (0 a: Type) -> (1 n: Nat) -> (xs: Vect n a) -> Nat
len a 0 []             = ?len_rhs_0
len a (S n') (x :: xs) = ?len_rhs_1
\end{idrblk}
in which \verb|n'| in the context of \verb|?len_rhs_1| is an unrestricted term.
With this property, we may introduce the \textit{unrestricted modality} from linear logic \cite{girard1987linear} as a type as follows:
\begin{idrblk}
record (!*) (a: Type) where
  constructor MkBang
  unrestricted: a
\end{idrblk}
with what a term of type \verb|!* a| with multiplicity 1 will be used linearly, while it can be consumed by pattern matching and produce an unrestricted term of type \verb|a|.

\begin{figure}[!t]
    \centering
    \includegraphics[width=\linewidth]{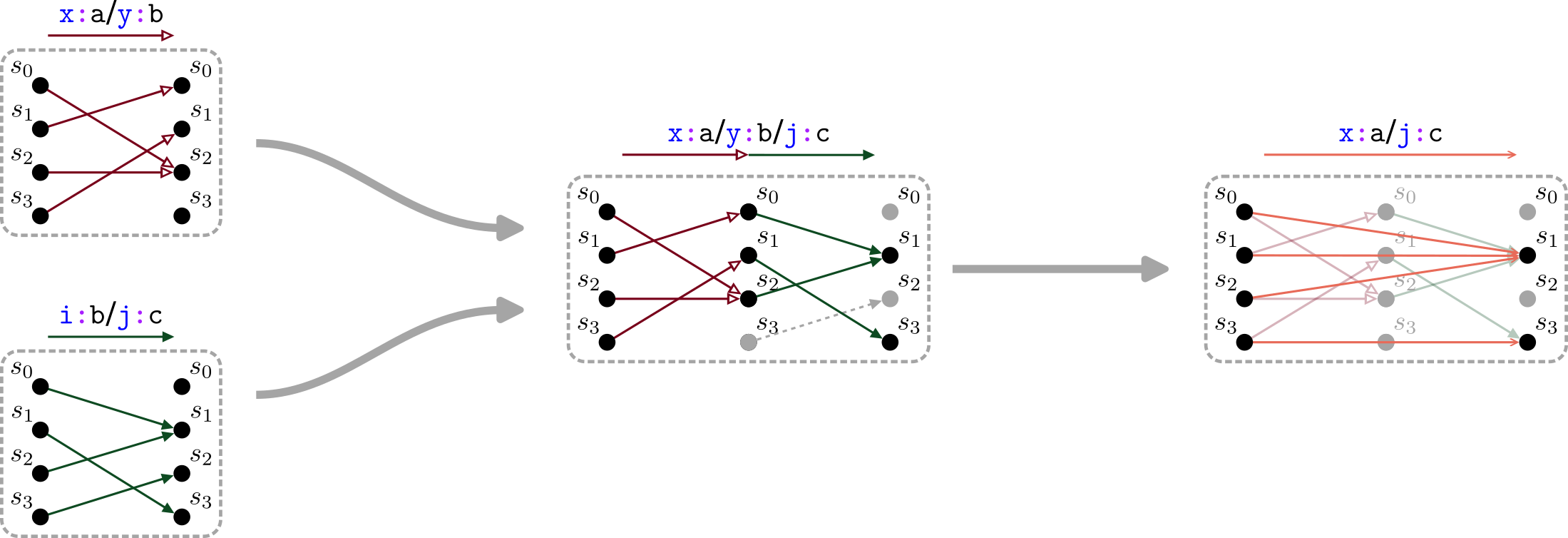}
    \caption{The illustration of the sequential composition of Mealy machines characterised by functions of type \protect\Verb|(x: a) -> (st: s) -> (b, s)| and \protect\Verb|(i: b) -> (st: s) -> (c, s)|, respectively, which share a state space with four states \(s_1 \dots s_4\). The result of composition is another function of type \protect\Verb|(x: a) -> (st: s) -> (c, s)|.}
    \label{fig:seq-comp-mealy}
\end{figure}

Finally, we use an example to illustrate how enforcing linearity allows components to be properly composed.
In this example, we consider the \textit{sequential composition} of processes sharing the same state space.
A process is considered reactive and stateful and, hence, is specified as a Mealy machine that is characterised by an initial state and a function that maps its input and current state to its output and next state.
The function characterising a Mealy machine, whose input, output and state are respectively of type \verb|a|, \verb|b| and \verb|s|, is of the (unrestricted) type \verb|(x: a) -> (st: s) -> (b, s)|.
The sequential composition of two such processes is then defined by a higher-order function that consumes their characteristic functions and produces the function characterising the composed process.
The expected behaviour of this composition is illustrated in \Cref{fig:seq-comp-mealy}.
The type signature and the correct implementation of the sequential composition can be given straightforwardly as follows:
\begin{idrblk}
seq: (f: (x: a) -> (st: s) -> (b, s))
  -> (g: (i: b) -> (st: s) -> (c, s))
  -> (x: a) -> (st: s) -> (c, s)
seq f g x st = let (y, st') = f x st in g y st'
\end{idrblk}

However, in the context of system design, the design space implied by the type of \verb|seq| consists of \textit{incorrect} implementations that \textit{cannot} be ruled out by a dependent-type-based specification.
For instance, the following is another possible implementation, named \verb|seq'| for distinguishing, which is obtained by inlining \verb|(y, st') = f x st| in \verb|seq|:
\begin{idrblk}
seq' f g x st = g (fst $ f x st) (snd $ f x st)
\end{idrblk}
whose equivalence to \verb|seq| is asserted by the following function:
\begin{idrblk}
seq_eq: (f: (x: a) -> (st: s) -> (b, s)) 
     -> (g: (i: b) -> (st: s) -> (c, s))
     -> (x: a) -> (st: s)
     -> seq f g x st = seq' f g x st
seq_eq f g x st with (f x st)
  seq_eq f g x st | (y, st') = Refl
\end{idrblk}
The behaviour of \verb|seq'|, however, is not always identical to \verb|seq|.
Counterexamples can be observed when composing functions that take the \textit{identifier} of the state, e.g. a pointer in C, instead of the value of the state as their input.
As an example, we may consider programs (functions) \verb|f| and \verb|g| specified in C presented in \Cref{fig:idr-glue-u-c}, which are introduced into Idris2 with proper types through the foreign function interface (FFI) as shown in \Cref{fig:idr-glue-u-idr}.
For any input \verb|x: Int| and state \verb|st: Ptr Int|, the output value returned by \verb|seq f g x st| will always be 0 while \verb|seq' f g x st| will return the value of \verb|x|.
It is because of that, in \verb|seq'|, \verb|f| is invoked twice, and each invocation updates the state.

\begin{figure}[t!]
    \centering
\begin{subfigure}[t]{0.5\linewidth}
        \begin{codeblksmall}
int f (int x, void* st){       
  int *st_ptr = st;              
  *st_ptr += x;                  
  return *st_ptr;
}

int g (int i, void* st){
  int *st_ptr = st;
  *st_ptr -= i;
  return *st_ptr;
}  
        \end{codeblksmall}
        \caption{}
        \label{fig:idr-glue-u-c}
    \end{subfigure}%
~ 
    \begin{subfigure}[t]{0.5\linewidth}
        \begin{codeblksmall}
%foreign "C:f,some_lib"
f': Int -> Ptr Int -> Int

f: (x: Int) -> (st: Ptr Int) 
 -> (Int, Ptr Int)
f x st = (f' x st, st)

%foreign "C:g,some_lib"
g': Int -> Ptr Int -> Int

g: (i: Int) -> (st: Ptr Int) 
 -> (Int, Ptr Int)
g i st = (g' i st, st)
        \end{codeblksmall}
        \caption{}
        \label{fig:idr-glue-u-idr}
    \end{subfigure}
    \caption{C functions that take pointers of states as input (a) and the corresponding Idris2 declarations that introduce these functions by the foreign function interface (FFI) (b).}
    \label{fig:idr-glue-u}
\end{figure}

Our example can be simply fixed by modifying \verb|f| and \verb|g| so that they operate on the value of the state, but it only works because the state in the example is an integer that can be passed without extra cost.
For large states, e.g., arrays, passing identifiers should, at least, be an option. 
In such cases, leveraging \textit{linearity} enables us to properly type these processes and their sequential composition.
To avoid over-constraining a process, we may consider restricting a process with linearity \textit{only} on how it manipulates the state.
That is to say, the invocation of a process is expected to \textit{replace} the current state with a new state (linearity) while its input and output can be used \textit{unrestrictedly}.
This constrain then leads us to the following type signature of a process:
\begin{idrblk}
proc: (x: a) -> (1 st: s) -> LC b s
\end{idrblk}
in which \verb|LC b s| is the product of an unrestricted value of type \verb|b| and a linear state of type \verb|s|, which is defined as:
\begin{idrblk}
data LC: Type -> Type -> Type where
  (#): (io: a) -> (1 st: s) -> LC a s
\end{idrblk}
Consequently, the type signature of the sequential composition can then be given as:
\begin{idrblk}
seq'': (1 f: (x: a) -> (1 st: s) -> LC b s)
    -> (1 g: (i: b) -> (1 st: s) -> LC c s)
    -> (x: a) -> (1 st: s) -> LC c s
\end{idrblk}
in which \verb|f| and \verb|g| are also bound with multiplicity 1, i.e., besides the state, each process should also be used linearly.
We can then verify that the type of \verb|seq''| \textit{rejects} the second implementation (\verb|seq'|) of sequential composition by type-checking the following:
\begin{idrblk}
seq'' f g x st = g ?g_fst (snd $ f x st)
\end{idrblk}
which gives us:
\begin{idrblk}
 `--             x : a
             0  st : s
              0  g : b -> (1 _ : s) -> LC c s
              0  f : a -> (1 _ : s) -> LC b s
     -----------------------------------------
      "Main.g_fst" : b 
\end{idrblk}
indicating that neither \verb|f| nor \verb|st| should be used at the position \verb|?g_fst|.
\subsection{The Tagless Final Approach}
\label{sec:preliminary:tf}

The introduction to QTT in the previous subsection sketched how QTT is planned to be used in a design framework.
In this section, we introduce the tagless final approach \cite{carette2009finally} with higher-order abstract syntax (HOAS) \cite{pfenning1988higher}, which enables an EDSL to fully leverage the infrastructure of the meta-language. 
To this end, a design process can benefit from both QTT natively implemented by Idris2 and the expressiveness facilitated by introducing DSLs.

In the tagless final approach, an EDSL is introduced by declaring interface(s), better known as \textit{type classes} in Haskell \cite{jones1993system,jones1997type,hudak2007history}, parametrised by type constructors\footnote{Type constructors can be considered as type-level functions that produce types by their application, e.g, the \Verb|Vect| referred in \Cref{judg:vec-len}.}.
The interface(s) that defines the EDSL is recognised as the \textit{symantics} (\textbf{sy}ntax + se\textbf{mantics}) of the EDSL.
And an implementation of the symantics on a concrete type constructor forms an interpretation of the EDSL.
Considering that an interpretation of a term in the EDSL models an aspect of a system, the tagless final approach enables us to write well-typed terms in the EDSL in a form that is \textit{polymorphic} on all interpretations and, hence, on all aspects.

\begin{figure}[t!]
    \centering
\begin{subfigure}{0.9\linewidth}
        \begin{codeblksmall}
interface Sym (repr: Type -> Type) where
  fn: (repr a -> repr b) -> repr (a -> b)
  app: repr (a -> b) -> repr a -> repr b
  add: repr Int -> repr Int -> repr Int
  cmp: repr Int -> repr Int -> repr Bool
  const: Int -> repr Int
        \end{codeblksmall}
        \caption{The symantics of the EDSL introduced as an interface in Idris2 parameterised by the abstract type constructor \protect\Verb|repr: Type -> Type|}
        \label{fig:tf-example-symantics}
    \end{subfigure}%
\vskip 1em
    \begin{subfigure}{\linewidth}
        \begin{codeblksmall}
record Sem a where
  constructor MkSem
  sem: a
        
Sym Sem where
  fn f    = MkSem (sem . f . MkSem )
  app f e = MkSem ((sem f) (sem e))
  add x y = MkSem ((sem x) + (sem y))
  cmp x y = MkSem ((sem x) == (sem y))
  const x = MkSem x
        \end{codeblksmall}
        \caption{The interpreter of the EDSL that maps terms written by constructors illustrated in \Cref{fig:tf-example-symantics} to terms in Idris2.}
        \label{fig:tf-example-interp}
    \end{subfigure}
    \caption{The symantics and interpreter of a small EDSL that supports function definition, application, integer addition, comparison and declaring constants.}
    \label{fig:tf-example}
\end{figure}

As an example, we may consider a small EDSL, whose symantics is illustrated in \Cref{fig:tf-example-symantics}, that supports function definition (\verb|fn|), application (\verb|app|), integer addition (\verb|add|) and comparison (\verb|cmp|), as well as constants declaration (\verb|const|).
In this semantics, \verb|repr: Type -> Type| is the \textit{abstract} type constructor parametrising the interface, which allows us to write terms that are interpretation-independent.
For instance, the following is \textit{a function in the EDSL} that maps integers to booleans:
\begin{idrblk}
term1: {repr:_} -> (Sym repr) => repr (Int -> Bool)      
term1 = fn (\x => cmp x (add x (const 1)))
\end{idrblk}
whose type signature can be read as: in all interpreters (\verb|{repr: _}|) of the EDSL (\verb|(Sym repr)|), we can construct a term of the type \verb|Int -> Bool| (\verb|repr (Int -> Bool)|).
Meanwhile, the implementation of \verb|term1| proves the claim by constructing a term of the type \verb|repr (Int -> Bool)|.
Note that \verb|term1|'s input variable is bound to the name \verb|x| \textit{in the meta-language}, which is the application of HOAS.

From the logic framework point of view  \cite{harper1993framework}, specifying an EDSL's symantics \textit{implicitly} defines typing rules desired by the EDSL.
That is, we can regard a type signature defined in the symantics as a derivation rule in the type system by regarding an arrow (\verb|->|) in the meta-language as a \textit{syntactic entailment}\footnote{By syntactic entailment, an arrow is translated to either the turnstile ($\vdash$) or the horizontal bar in a derivation rule.} in the type system.
For instance, the \verb|fn| constructor in \Cref{fig:tf-example-symantics} can be interpreted as:
\begin{center}
\begin{prooftree}
     \hypo{ x : \mathtt{repr\:a} \vdash y:\mathtt{repr\:b}}
    \infer{1}[(fn)]{ \vdash \lambda x. y : \mathtt{repr \: (a \to b)}}
\end{prooftree}    
\end{center}
which is, in essence, the ($\to\text{Intro}$) rule in the EDSL.
From this point of view, we can type check (proof) the judgement that \verb|term1: repr (Int -> Bool)| by the following derivation tree
\begin{center}
\scalebox{0.85}{
\begin{prooftree}
    \hypo{x : \mathtt{repr\:Int} \vdash x : \mathtt{repr\:Int}}
    \hypo{x : \mathtt{repr\:Int} \vdash x : \mathtt{repr\:Int}}
    \hypo{\vdash (\mathtt{const} \: 1) : \mathtt{repr\:Int}}
    \infer{2}[(add)]{x : \mathtt{repr\:Int} \vdash (\mathtt{add} \: x \: (\mathtt{const} \: 1)): \mathtt{repr \: \mathtt{Int}}}
     \infer{2}[(cmp)]{ x : \mathtt{repr\:Int} \vdash \mathtt{cmp} \: x \: (\mathtt{add} \: x \: (\mathtt{const} \: 1)):\mathtt{repr\:Bool}}
    \infer{1}[(fn)]{ \vdash \lambda x. \mathtt{cmp} \: x \: (\mathtt{add} \: x \: (\mathtt{const} \: 1)): \mathtt{repr \: (\mathtt{Int} \to \mathtt{Bool})}}
\end{prooftree}}
\end{center}
in which \(x : \mathtt{repr\:Int} \vdash x : \mathtt{repr\:Int}\) hold because of the existence of the \textit{identity function} in the meta-language.
To this end, we can benefit from the tagless final approach to easily define the type system of an EDSL based on the meta-language's type system.
Furthermore, since the context of a judgement in the EDSL's type system can be explicitly managed in the meta-language, the expressiveness of the EDSL's type system is not limited by the meta-language's type system.
For example, we can embed a language with linear types into vanilla Haskell \cite{polakow2015embedding}.

An \textit{interpreter} for an EDSL embedded by the tagless final approach \textit{specialises} the EDSL's symantics to specific semantics by implementing the corresponding interface(s) with a \textit{concrete} type constructor.
For instance, \Cref{fig:tf-example-interp} illustrates an implementation of the symantics \verb|Sym| presented in \Cref{fig:tf-example-symantics} with the type constructor \verb|Sem: Type -> Type|, which, as a record, is defined by a \textit{data constructor} \verb|MkSem: a -> Sem a| that send a term in Idris2 of type \verb|a| to a term of type \verb|Sem a| and a \textit{projection} \verb|sem: Sem a -> a| which, in this case, is the inverse of \verb|MkSem|.
By the implementation presented in \Cref{fig:tf-example-interp}, the projection \verb|sem| forms the interpreter, which, in essence, denotes terms in the EDSL by terms in Idris2.
For instance, the implementation of \verb|add| constructor suggests the following:
\begin{idrblk}
sem (add x y) = (sem x) + (sem y)
\end{idrblk}
in which (\verb|+|) is the addition in Idris2.
Similarly, we can interpret \verb|term1| as an Idris2 function by simply writing \verb|sem term1|, which is of the type \verb|Int -> Bool|.
Note that an interpreter defined by a concrete type constructor \verb|I| need not be exactly of the type \verb|I a -> a|.
Instead, it can be \textit{indexed} by other variables, which allows us to define interpreters that interpret terms in a context-sensitive manner.
To this end, interpretations can also be optimisation steps, e.g., we may make an interpreter that identifies and eliminates shared expressions \cite{kiselyov2011implementing}.

\section{The SynQ Language}
\label{sec:sym}
This section presents the SynQ language.
We will start with a general discussion about how synchronous systems are modelled in SynQ.
And then introduce the symantics that defines SynQ.
\subsection{Component-based Specifications of Synchronous Systems}
\label{sec:sym:cbd}
Synchronous systems are reactive systems obeying the \textit{perfect synchrony hypothesis} that \textquote{reactions are instantaneous so that activations and productions of output are synchronous, as if programs were executed on an infinitely fast machine} \cite{boussinot1991esterel}.
In other words, a system's synchronous model abstracts out (physical) time computations consumed.
Meanwhile, the order between a system's input events, and hence between its output events/firings, is preserved, implying the existence of a (possibly aperiodic) \textit{global clock} in a system that synchronises the entire system. 

\begin{figure*}[t]
    \centering
    \includegraphics[width=\linewidth]{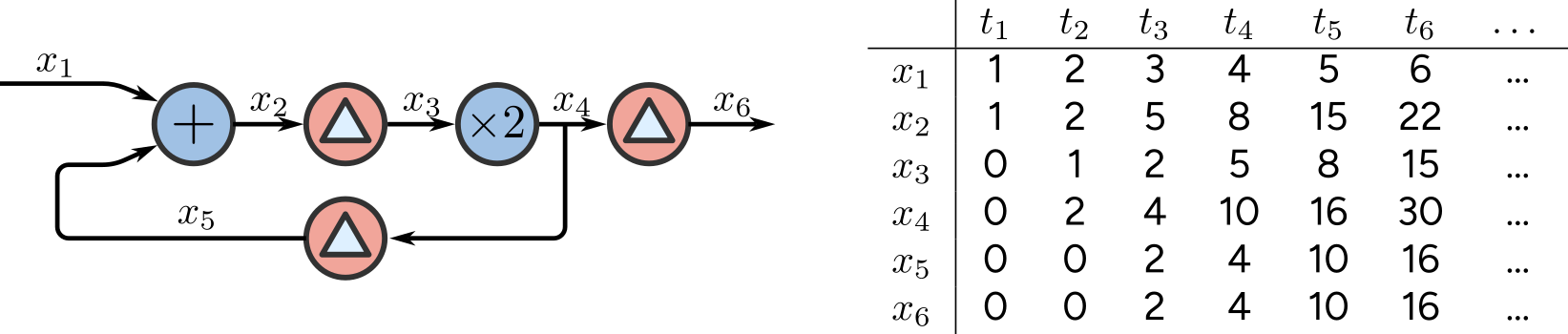}
    \caption{An exemplified synchronous system (left) illustrated as a network consists of combinational operators (blue) and unit delays (red); and streams' value at each clock event by assuming all unit delays have initial value 0.}
    \label{fig:sync-example-1}
\end{figure*}

Synchronous systems' behaviour can be specified by the dataflow model as a set of equations over \textit{synchronous isomorphic} (cf. Section 3 of \cite{gammie2013synchronous}) streams (sequences of events, dataflow) \cite{gilles1974semantics,pilaud1987lustre} that are mutually recursive.
For instance, the following set of equations defines a synchronous system:
\begin{equation*}
   \begin{aligned}[t]
        & x_2 = x_1 + x_5 \\
        & x_3 = D\: x_2\\
        & x_4 = x_3 \times 2
    \end{aligned}
\quad\quad\quad\quad
    \begin{aligned}[t]
        & x_5 = D \: x_4 \\
        & x_6 = D \: x_4
    \end{aligned}
\end{equation*}
which can be directly interpreted as a netlist and visualised as a network illustrated in \Cref{fig:sync-example-1}.
In this set of equations, $x_{1-6}$ are streams indexed by the clock event, e.g., $x_1[k]$ refers to the value of $x_1$ at the $k$-th clock event, and the constant 2 represents a constant stream of value 2.
Operators \((+)\) and \((\times)\) indicate \textit{point-wise} (combinational) operations over streams and \(D\) is the primitive of unit delay such that \(x[k-1] = (D \: x) [k]\) with a pre-defined \((D \: x) [0]\).

The equation-based specification of synchronous models is formal and declarative.
However, this specification methodology lacks compositionality and, hence, is challenging to use for large and complex systems building with smaller and simpler components.
That is, with this specification methodology, a desired property of the system, e.g., causality\footnote{By causality, we mean here a system's output at the \(k\)-th clock event can be determined by the system's input from time 0 to \(k\).}, may not be guaranteed by composing sub-systems that have the property.
For example, we may consider the following top-level specification of a synchronous system:
\begin{center}
\begin{minipage}{0.4\linewidth}
\centering
    \begin{equation*}
        x_2 = S \: x_1 \: x_2
    \end{equation*}
\end{minipage}
\begin{minipage}{0.4\linewidth}
\centering
    \includegraphics{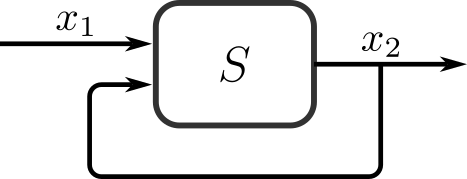}
\end{minipage}
\end{center}
in which \(S\) is an arbitrary sub-system.
It is clear that the system's causality is not guaranteed by ensuring the causality of the sub-system \(S\). Instead, ad-hoc reasoning is required for each instantiation of \(S\).

SynQ adopts a component-based design framework to achieve compositionality, in which, instead of binding streams to names (variables) and specifying a synchronous system as a set of equations that captures relations between these variables, systems (components) are constructed by applying \textit{glue components} on components (sub-systems).
In other words, systems in SynQ are constructed by composing sub-systems but not \textit{wiring} ports from different sub-systems.
To this end, compositionality in SynQ is achieved by carefully selecting the set of glue and atomic components, which will be described later in this section.

To correctly handle \textit{feedback loops} during composing components, components with feedback loops are specified in the form that:
\begin{itemize}
    \item the occurrence of combinational operations and delays are strictly limited to the forward (from a system's input to its output) and backward path(s), respectively; and
    \item there exists \textit{exactly one} unit delay per backward path.
\end{itemize}
In this way, we can simply specify the \textit{existence} of feedback loops without saying what a backward path consists of.
Note that a forwarded unit delay can still be introduced as a component as follows:
\begin{center}
    \includegraphics{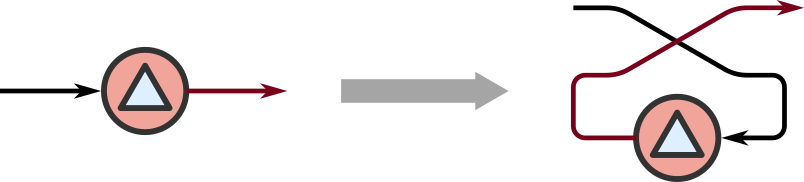}
\end{center}
Meanwhile, even though each backward path is restricted to have exactly one unit delay, SynQ still has an expressiveness that is similar to Lustre, in which only zero-delay loops are forbidden \cite{pilaud1987lustre}.
This can be illustrated by the following example of how a feedback loop with two delays is specified in the desired form.
\begin{center}
    \includegraphics[width=\linewidth]{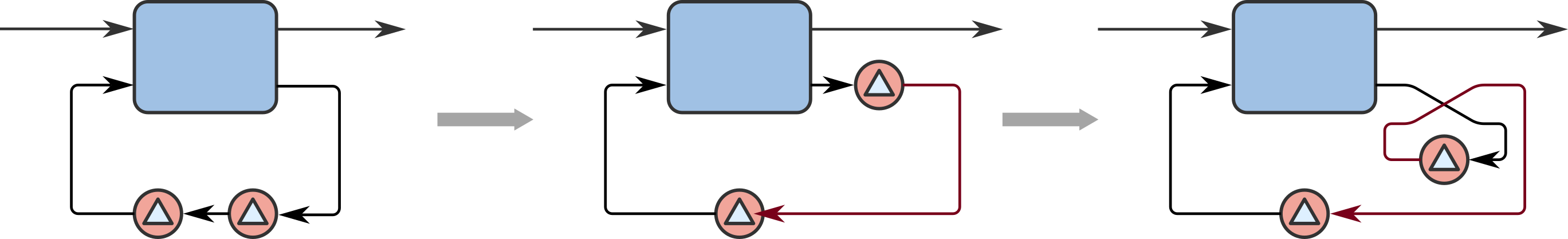}
\end{center}
Besides the restricted form a feedback loop must have, SynQ also leverages QTT to restrict compositions of components with feedback loops by \textit{typing each backward path with multiplicity 1 (linearity)}, which will be discussed later when the symantics and interpreters of SynQ are introduced.
\subsection{The Symantics of SynQ}
\label{sec:sym:sym}

We are now ready to introduce SynQ by its symantics.
To avoid over-constraining the expressiveness of SynQ, the symantics of SynQ are separated into two layers, namely the \textit{combinational} and \textit{sequential} layer, respectively.
Constructors in the combinational layer aim to design ``loop-free'' systems efficiently and correctly by fully leveraging its meta-language Idris2 to provide structural descriptions of systems.
Consequently, linearity is not enforced in the combinational layer.
Components with feedback loops and, generally, states that need to be processed sequentially are introduced and composed by constructors in the sequential layer.
Linearity is leveraged in this layer to guarantee that only ``proper'' composition can be made, with a similar intuition as the example presented in \Cref{sec:preliminary:qtt}.
To support a component-based design framework, within each layer, constructors are further classified into two sets that respectively correspond to \textit{glue} and \textit{atomic components}.
Later in \Cref{sec:interpreter}, we will present that our classification of glue and atomic components is aligned with the intuition proposed in \cite{sifakis2015system}.

\subsubsection{The Combinational Layer}
\label{sec:sym:comb}

\begin{figure}[t!]
    \centering
\begin{subfigure}{0.9\linewidth}
        \begin{codeblksmall}
interface Comb (comb: Type -> Type -> Type) where
  lam  : {auto aIsSig: Sig a} -> {auto bIsSig: Sig b}
      -> (comb () a -> comb () b) -> comb a b
  app  : {auto aIsSig: Sig a} -> {auto bIsSig: Sig b}
      -> comb a b -> comb () a -> comb () b
  prod : {auto aIsSig: Sig a} -> {auto bIsSig: Sig b}
      -> comb () a -> comb () b -> comb () (a, b)
  proj1: {auto aIsSig: Sig a} -> {auto bIsSig: Sig b}
      -> comb () (a, b) -> comb () a
  proj2: {auto aIsSig: Sig a} -> {auto bIsSig: Sig b}
      -> comb () (a, b) -> comb () b
  unit : comb () ()
        \end{codeblksmall}
        \caption{The interface that defines the symantics of combinational glue components of SynQ.}
        \label{fig:synq-symantics-comb-glue}
    \end{subfigure}%
\vskip 1em
    \begin{subfigure}{\linewidth}
        \begin{codeblksmall}
interface Primitive (comb: Type -> Type -> Type) where
  const: {n: Nat} -> BitVec n -> comb () (BitVec n)  
  add   : {n: Nat} -> comb () (BitVec n) 
       -> comb () (BitVec n) -> comb () (BitVec $ n + 1)
  concat: {m:_} -> {n:_} 
       -> comb () (BitVec m) -> comb () (BitVec n) 
       -> comb () (BitVec $ m + n)
  ...
  slice : {n: Nat} -> (lower: Nat) -> (upper: Nat) 
       -> {auto prf_upper: LTE upper n}
       -> {auto prf_lower: LTE lower upper}
       -> comb () (BitVec n) 
       -> comb () (BitVec $ minus upper lower)  
  ...
        \end{codeblksmall}
        \caption{The interface that defines the symantics of combinational atomic components of SynQ}
        \label{fig:synq-symantics-comb-atomic}
    \end{subfigure}
    \caption{The symantics of the combinational layer of SynQ}
    \label{fig:synq-symantics-comb}
\end{figure}

The symantics of the combinational layer is illustrated in \Cref{fig:synq-symantics-comb}, in which only a few representative atomic components (primitives) are presented.
A component (term) in the combinational layer is represented by the (abstract) type constructor \verb|comb: Type -> Type -> Type|, which suggests that a component of the type  \verb|comb a b| is parametrised by two types, \verb|a| and \verb|b|, respectively.
Even though no interpretations of components can be made by solely giving the symantics, we may still interpret types parametrising a component as types of a component's input and output because they restrict how components are composed.
In this way, a component of the type  \verb|comb a b| is considered a component with input and output of type \verb|a| and \verb|b|, respectively.
\verb|()| (the unit type in the meta-language) indicates the absence of the corresponding port.
Consequently, \verb|comb () a| can be interpreted as a signal (constant component) of type \verb|a| whose value is not dependent on other signals.

Glue and atomic components are parametrised differently.
Glue components are parametrised by type variables, i.e.,  \verb|a|, \verb|b|, \verb|c| etc., while atomic components' input/output can only be bit-vectors.
This is because glue components are used to make structural descriptions of systems and, hence, need to be generic for all components whose ports are matched.
Atomic components, on the other hand, are basic building blocks that are used as black boxes.
They are, hence, parametrised by (polymorphic to) the length of bit-vectors only so that they can later be specialised to primitives of common data types, e.g., integer, floating point, boolean, etc.

Besides using types that are parametrising components to constrain the composition of components, constructors consisting in the symantics also employ \textit{type-based contracts} to further guarantee the correct application of constructors.
These contracts are wrapped in curly brackets and annotated with \verb|auto|, which means that they are expected to be implicitly reasoned by the compiler.
For an atomic component, a contract is used so that its input(s) satisfies some requirements.
For instance, \verb|{auto prf_upper: LTE upper n}| ($\texttt{upper} \leq n$) in \verb|slice| suggests that the highest position (\verb|upper|) of a slice of a bit-vector do not exceed the length of the vector.

Glue components employ a common contract, which is formed as a type constructor \verb|Sig: Type -> Type| that is defined as follows.
\begin{idrblk}
data Sig: Type -> Type where
  U : Sig Unit
  BV: {n:_} -> Sig $ BitVec n
  P : Sig a -> Sig b -> Sig (a, b)
\end{idrblk}
By this definition, \verb|Sig| forms a \textit{predicate over types}, which inductively defines the following subset of types:
\begin{itemize}
    \item the unit type (\verb|Unit|/\verb|()|) belongs to the subset (\verb|Sig Unit|);
    \item the type of bit-vectors with arbitrary length belongs to the subset (\verb|Sig $ BitVec n|); and
    \item if \verb|a| and \verb|b| belong to the subset, then so does their product \verb|(a, b)| (\verb|Sig (a, b)|).
\end{itemize}
The use of \verb|Sig| allows only components with specific input/output types to be constructed in SynQ and consequently restricts the expressiveness of SynQ.
With this restriction, we can guarantee that \textit{all systems that can be constructed by the combinational layer of SynQ are \textit{reactive}\footnote{In other words, the computation on an input event is guaranteed to be terminated, which is because of that the language is then equivalent to first-order simply typed lambda calculus.}}.
Otherwise, even though our intention is to specify loop-free systems in the combinational layer, we can still specify a term of the type:
\verb|comb (comb a b) c| which may entail infinitely many applications of another component of type \verb|comb a b|.

Glue components specified in \Cref{fig:synq-symantics-comb-glue} can be categorised into two sets.
One of them allows us to introduce and use components (\verb|lam| and \verb|app|), which is similar to \verb|fn| and \verb|app| presented in \Cref{fig:tf-example-symantics} by regarding the prefix operator \verb|comp| as the infix operator \verb|(->)|.
And the other enables us to pack and unpack \textit{signals} (\verb|prod|, \verb|proj1|, \verb|proj2| and \verb|unit|).
As we discussed in \Cref{sec:preliminary:tf}, formally, by specifying the symantics, we are giving typing rules of terms in SynQ.
But we could also informally visualise a glue component, e.g., \verb|app|, as follows:
\begin{center}
    \includegraphics{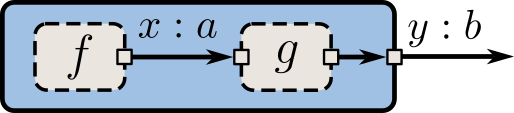}
\end{center}
in which dashed boxes $f$ and $g$ represent components to be glued, and the out-most box indicates the \textit{shape} (type) of the result component.
This visualisation is aligned with the typing rule that \textit{we can obtain \Verb|app f g: comb () b| if we have \Verb|f: comb () a| and \Verb|g: comb a b|}.
Note that the constructor \verb|lam| is not a conventional component. 
Instead, it should be considered as a process in the meta-language that transforms a signal parametrised by another to a component, as shown below.
\begin{center}
    \includegraphics{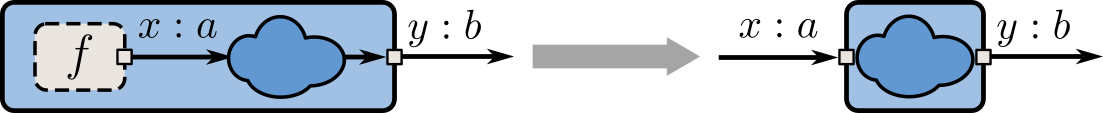}
\end{center}

Most of the glue components in \Cref{fig:synq-symantics-comb-glue} result in \textit{signals}.
However, this set of glue components is adequate for constructing glues for components with input/output.
For instance, a sequential composition of two components can be given as:
\begin{center}
\begin{minipage}[t]{0.55\linewidth}
\centering
\begin{codeblksmall}
(<<): {comb:_} -> (Comb comb)
   => {auto aIsSig: Sig a} 
   -> {auto bIsSig: Sig b}
   -> {auto cIsSig: Sig c}
   -> comb b c -> comb a b -> comb a c
(<<) g f = lam $ \x => app g $ app f x
\end{codeblksmall}
\end{minipage}
\begin{minipage}[t]{0.4\linewidth}
\centering
    \includegraphics[width=0.8\linewidth]{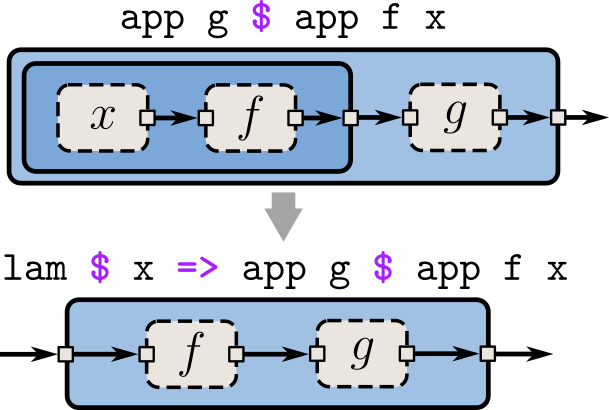}
\end{minipage}
\end{center}
Meanwhile, the parallel composition of two components can be specified as (implicit arguments are omitted):
\begin{idrblk}
(<>) : ... -> comb a b -> comb c d -> comb (a, c) (b, d)
(<>) f g = lam $ \xs => prod (app f $ proj1 xs) (app g $ proj2 xs)
\end{idrblk}

\begin{figure}[t!]
    \centering
\begin{subfigure}{\linewidth}
        \begin{codeblksmall}
reduce: {comb:_} -> (Comb comb)
     => {auto prf1: Sig a} -> {auto prf2: All (OfType a) as}
     -> (f: comb (a, a) a) -> comb as a
reduce {prf2 = (AllU p)}     f = rewrite sym $ p in lam id
reduce {prf2 = (AllP p1 p2)} f = f << (reduce f) <> (reduce f)
        \end{codeblksmall}
        \caption{}
        \label{fig:reduce-def}
    \end{subfigure}%
\vskip 1em
    \begin{subfigure}{\linewidth}
    \centering
\includegraphics{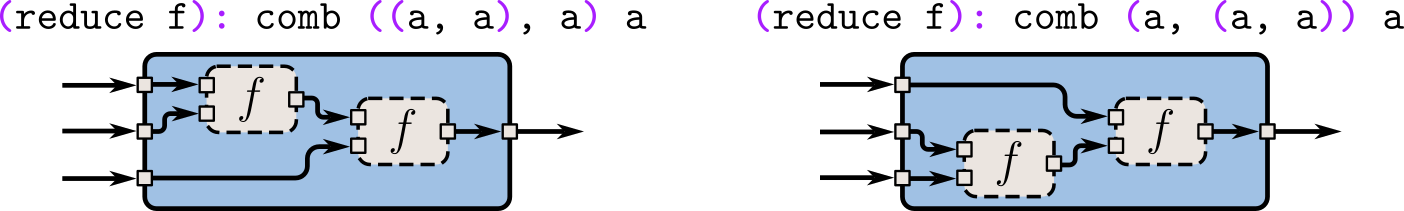}
        \caption{}
        \label{fig:reduce-viz}
    \end{subfigure}
    \caption{The specification of the \protect\Verb|reduce| glue (a) and the visualisation of applying the glue in contexts where input signals are packed differently (b). Note that in the visualisation, we did not distinguish how signals are packed due to synchronous isomorphism.}
    \label{fig:reduce}
\end{figure}

Besides these simple glues, we can also specify more complex glues in which the structure of a system is \textit{recursively defined} or \textit{determined by the context}.
As an example, we can construct a glue of \textit{reducer}, namely \verb|reduce| as defined in \Cref{fig:reduce-def}, which uses a component with two inputs, namely \verb|f: comb (a, a) a|, to reduce a set of signals of type \verb|a|, which is given by a (packed) signal of the type \verb|comb () (a, a, ..., a)|, to a signal of the type \verb|comb () a|.
Furthermore, as illustrated in \Cref{fig:reduce-viz}, the reducer applies \verb|f| in the order of how input signals are packed.
That is, applying \verb|reduce f| on a signal of the type \verb|comb () ((a, a), a)| (left) will apply \verb|f| on the first two inputs and then apply \verb|f| on the result of the first application and the last input.
Meanwhile, reducing a signal of type \verb|comb () (a, (a, a))| (right) will reduce the last two signals first and then the first signal and the result.

Two key enablers that allow us to construct \verb|reduce| are the induction in the meta-language and type constructor(s) \verb|All (OfType a) as|.
The induction is defined in the case:
\begin{idrblk}
reduce {prf2 = (AllP p1 p2)} f = f << (reduce f) <> (reduce f)   
\end{idrblk}
which is a statement in Idris2 instead of SynQ and, hence, is not limited by the expressiveness of SynQ.
The type constructor \verb|All (OfType a) as| is used to contract the type variable \verb|as| so that \verb|as| is the set of types that consists of only elements of type \verb|a| packed in different ways, which is defined as follows.
\begin{center}
\begin{minipage}[t]{0.45\linewidth}
\centering
\begin{codeblksmall}
OfType: Type -> Type -> Type
OfType x y = x = y


\end{codeblksmall}
\end{minipage}
\begin{minipage}[t]{0.54\linewidth}
\centering
\begin{codeblksmall}
data All: (pred: Type -> Type) 
       -> Type -> Type where
  AllU: {0 pred: _} -> {x: Type} 
     -> (pred x) -> All pred x
  AllP: {0 pred: _} -> All pred a 
     -> All pred b -> All pred (a, b)
\end{codeblksmall}
\end{minipage}
\end{center}
With this type constructor, the result type (\verb|comb as a|) of applying \verb|reduce| is polymorphic to all input signals that are homogeneous.
Meanwhile, by pattern matching on the generated proof of \verb|All (OfType a) as| in the implementation of \verb|reduce|, different structures are inductively generated when \verb|reduce| is applied in a context.

\paragraph{Remark} 
The implementation of \verb|reduce| reflects the trade-off of expressiveness taken by the design of SynQ.
Specifically, the expressiveness consideration in SynQ is twofold, including the expressiveness of SynQ and the expressiveness that can be achieved by SynQ and its meta-language Idris2.
The former is heavily restricted by the usage of the type system, e.g., by the \verb|Sig| predicate discussed earlier, so that systems can have deterministic behaviour that can be reasoned.
Meanwhile, extra expressiveness is obtained by the fact that we can write regular Idris2 programs, which are referred to as \textit{macros}, that generate terms in SynQ.
To this end, we can have an expressive \textit{design process} that produces \textit{correct} results only.
This idea can be observed in the literature about multi-level languages and staging \cite{nielson1992two,taha1997multi,turner2004total,megacz2011hardware}.

\subsubsection{The Sequential Layer}
\label{sec:sym:seq}

\begin{figure}[t!]
    \centering
\begin{subfigure}{0.9\linewidth}
        \begin{codeblksmall}
interface Comb comb 
  => Seq (comb: Type -> Type -> Type)
         (seq: Type -> Type -> Type -> Type) | seq where
  abst: {auto aIsSig: Sig a} -> {auto bIsSig: Sig b}
     -> {auto sIsState: St s}
     -> (1 _: comb () a -> seq s () b) -> seq s a b
  pure: {auto aIsSig: Sig a} -> {auto bIsSig: Sig b}
     -> {auto sIsState: St s}
     -> comb a b -> seq s a b
  (=<<): {auto aIsSig: Sig a} -> {auto bIsSig: Sig b}
      -> {auto cIsSig: Sig c} -> {auto sIsState: St s} 
      -> (1 _: seq s b c) -> (1 _: seq s a b) 
      -> seq s a c 
  (<<<): {auto aIsSig: Sig a} -> {auto bIsSig: Sig b}
      -> {auto cIsSig: Sig c} -> {auto s1IsState: St s1}
      -> {auto s2IsState: St s2} 
      -> (1 _: seq s2 b c) -> (1 _: seq s1 a b)
      -> seq (LPair s1 s2) a c
        \end{codeblksmall}
        \caption{The interface that defines the symantics of sequential glue components of SynQ.}
        \label{fig:synq-symantics-seq-glue}
    \end{subfigure}%
\vskip 1em
    \begin{subfigure}{\linewidth}
        \begin{codeblksmall}
interface Comb comb 
  => Reg (comb: Type -> Type -> Type)
         (seq: Type -> Type -> Type -> Type) | seq where
  constructor MkReg
  1 get: {auto aIsSig: Sig a} -> {auto sIsState: St s}
      -> {auto similar: SameShape a s}
      -> seq s () a
  1 set: {auto aIsSig: Sig a} -> {auto sIsState: St s}
      -> {auto similar: SameShape a s}
      -> comb () a -> seq s () ()
        \end{codeblksmall}
        \caption{The interface that defines the symantics of sequential atomic components of SynQ}
        \label{fig:synq-symantics-seq-atomic}
    \end{subfigure}
    \caption{The symantics of the sequential layer of SynQ}
    \label{fig:synq-symantics-seq}
\end{figure}

The symantics of the sequential layer is illustrated in \Cref{fig:synq-symantics-seq}.
Compared to the symantics of the compositional layer in \Cref{fig:synq-symantics-comb}, interfaces defining the sequential layer are parametrised by two abstract type constructors, indicating that a term in the sequential layer, represented by the \verb|seq| constructor, may depend on a term in the compositional layer that is represented by the \verb|comb| constructor.
Currently, in SynQ, there is only one set of sequential atomic components (\verb|Reg| illustrated in \Cref{fig:synq-symantics-seq-atomic}) is employed, which allows us to retrieve (\verb|get|) and store (\verb|set|) (white-box) states.
With \verb|Reg|, SynQ is adequate for specifying synchronous systems.
On the other hand, our definition of sequential glue components in \Cref{fig:synq-symantics-seq-glue} is compatible with black/grey-box atomic components with implicit internal states.
Extra atomic components can be easily introduced based on the specification methodology in \cite{chen2024qtt}.

A sequential component is of the (abstract) type \verb|seq s a b|.
This type specifies that the component has an input of type \verb|a|, an output of type \verb|b| and a feedback loop with unit delay of type \verb|s|.
Similar to components in the combinational layer, types parametrising a sequential component are restricted by type-based predicates as well.
Specifically, types of the input/output of a sequential component are from the same subset of types that parametrise combinational components.
States, on the other hand, are typed by another subset of types that are specified by the \verb|St| type constructor defined as:
\begin{idrblk}
data St: Type -> Type where
  LU: St ()
  LV: {n: _} -> St $ !* (BitVec n)
  LP: (st1: St a) -> (st2: St b) 
   -> St $ LPair a b
\end{idrblk}
How \verb|St| is defined is very similar to the definition of \verb|Sig|. 
Two differences are that the type of bit-vector is wrapped by the \verb|!*| (the unrestricted modality introduced in \Cref{sec:preliminary:qtt}), and the product type is replaced by \verb|LPair|, in which all elements preserve linearity after pattern matching.
\verb|St| allows us to specify white-box states in a linear context such that a state itself is used linearly (for composition) while the value in the state can be used unrestrictedly (for computation).

Since types of states and types of inputs/outputs are two different subsets, the type constructor \verb|SameShape: Type -> Type -> Type| is employed by atomic components.
This type constructor is defined as follows
\begin{idrblk}
data SameShape: Type -> Type -> Type where
  U: SameShape () ()
  BV: SameShape a (!* a)
  P:  (prfa: SameShape a b)
   -> (prfb: SameShape c d)
   -> SameShape (a, c) (LPair b d)
\end{idrblk}
which forms a relation between signals characterised by \verb|Sig| and states characterised by \verb|St|.
By asserting this relation, atomic components are guaranteed to be able to get a well-typed signal from a state and set a state by a signal.

Linearity is used distinctly in glue and atomic components.
For glue components, linearity is mainly applied to the components to be glued so that states produced by/consumed from them can be properly handled and, hence, backward path(s) can be properly introduced.
For instance, the \verb|(=<<)| and \verb|(<<<)| glue guarantees that any of their interpretation respectively implements the following:
\begin{center}
\includegraphics{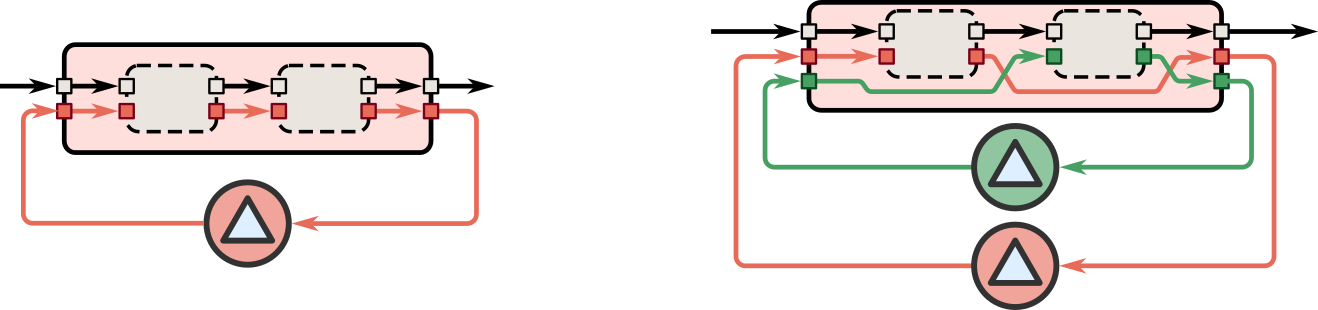}
\end{center}
in which coloured paths correspond to states.
An exception is the \verb|pure| constructor, which introduces a compositional component into a stateful context (alongside a feedback loop) and can be visualised as follows.
\begin{center}
\includegraphics{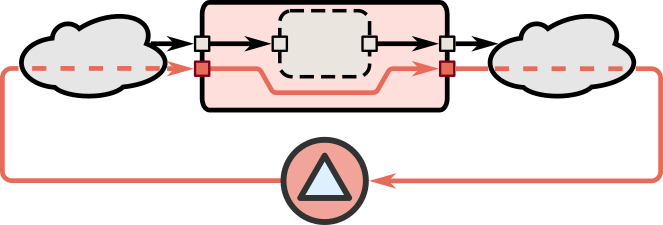}
\end{center}

For atomic components, the multiplicity 1 is denoted \textit{before} each constructor.
That is, constructors themselves are bound with multiplicity 1.
Consequently, these constructors shall be used exactly once in a linear context.
It allows us to precisely control the usage of atomic components, instead of how inputs of these components are used, in a context.

\begin{figure}[t!]
    \centering
\begin{subfigure}{\linewidth}
        \begin{codeblksmall}
scan: {comb: _} -> {seq: _} -> (Seq comb seq)
  => (1 reg: Reg comb seq) 
  -> ... -- implicit contracts on a, b, c, s
  -> {auto similar: SameShape c s}    
  -> (f: comb (a, c) (b, c)) -> seq s a b
scan (MkReg get set) f = 
  abst $ \x => 
        (abst $ \y => pure (proj1 y) =<< set (proj2 y)) -- 2
    =<< (pure (lam $ \y => app f $ prod x y) =<< get)   -- 1
        \end{codeblksmall}
        \caption{}
        \label{fig:scan-def}
    \end{subfigure}%
\vskip 1em
    \begin{subfigure}{\linewidth}
    \centering
\includegraphics{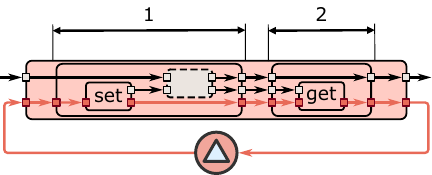}
        \caption{}
        \label{fig:scan-viz}
    \end{subfigure}
    \caption{The specification of the \protect\Verb|scan| glue (a) and its visualization (b).}
    \label{fig:scan}
\end{figure}

As a wrap-up example, here we consider the specification of a common pattern \verb|scan|.
The \verb|scan| glue can be seen as a reactive variant of \textit{fold} \cite{meijer1991functional,hutton1999fold} that converts a combinational component of the type \verb|comb (a, s) (b, s)| to a Mealy machine specified as a sequential component of the type \verb|seq s a b|.
Its specification in SynQ is illustrated in \Cref{fig:scan}.
With \verb|scan|, a forward delay can be simply given as:
\begin{idrblk}
dly: ... -> {_: SameShape a s} -> seq s a a
dly = scan (lam $ \x => prod (proj2 x) (proj1 x))
\end{idrblk}

Note that in \verb|scan|'s type signature, the symantics of sequential atomic components \verb|Reg| occurs \textit{after} the \verb|=>| and is bound with multiplicity 1.
In such a case, \verb|reg: Reg ...| is used as a linear resource instead of a set of typing rules, which can be used once for constructing \verb|scan| in the context.
This can be shown by inspecting the context formed by \verb|scan|:
\begin{idrblk}
`--                           f : comb (a, c) (b, c)
                        similar : SameShape c s
                               ...
                       sIsState : St s
                         1  reg : Reg comb seq
     ----------------------------------------------------------
                     "scan_rhs" : seq s a b
\end{idrblk}
Since constructors in \verb|Reg| are bound with linearity, each of them in this context can be used exactly once.
It showcases how resource management is facilitated by leveraging QTT in SynQ.

\section{Interpreters of SynQ}
\label{sec:interpreter}
The symantics abstractly defines SynQ.
That is, the symantics of a term (system model) forms the \textit{intersection} of this term's all possible aspects (interpretations), which need to be further interpreted to a concrete meaning.
As we have introduced earlier, an interpretation is made by implementing interfaces that define the symantics with a concrete type constructor.
The implementation is then referred to as an interpreter of SynQ.

This section illustrates three interpreters, which interpret SynQ terms to Idris2 functions, typed netlists and other EDSLs' symantics.
These interpreters, respectively, allow us to generate runnable programs, concrete intermediate representations, which can be used for optimisation and synthesisable HDL code generation, and abstract terms for further interpretations.
Further, they showcase three different ways of implementing interpreters.

\subsection{The Interpreter to Idris2 Functions}
\label{sec:interp:fn}

The interpreter maps SynQ terms to Iidris2 functions following exactly the same idea as the interpreter illustrated in \Cref{fig:tf-example-interp} discussed in \Cref{sec:preliminary:tf}.
This interpreter maps a term \textit{precisely} to a function in Idris2.
By precise, we mean here all types parameterising the term are interpreted by the function it mapped to.

The type constructor based on what the combinational layer of SynQ is interpreted is the following:
\begin{idrblk}
record Combinational a b where
  constructor MkComb
  runComb: a -> b
\end{idrblk}
Recall that a record is defined by projections from itself to terms of other types, \verb|runComb| is of the type \verb|Combinational a b -> (a -> b)|.
Hence, by properly implementing the \verb|Comb| and \verb|Primitive| interface with \verb|Combinational|, a term in SynQ of type \verb|comb a b| can be mapped to a function in Idris2 of type \verb|a -> b|, which makes \verb|runComb| the interpreter we expected in this section.

It is worth mentioning here that as a component-based design framework, components are composed based on their interfaces (types) instead of their implementations.
Hence, we can choose implementations of components that are not native Idris2 functions, which can be introduced by the elaborator \cite{christiansen2016elaborator} or FFIs (as illustrated in \Cref{fig:idr-glue-u}).
In the implementation of SynQ, combinational atomic components on bit-vectors are implemented in C and introduced by FFIs.
In this way, we can avoid the inefficiency caused by the inductively defined data types and obtain efficient runnable programs from this interpreter.
It also demonstrates how SynQ can be used with legacy components and/or reuse existing components as black boxes.

To interpret terms in the sequential layer, the following type constructor and its paired function are employed:
\begin{idrblk}
data Sequential: Type -> Type -> Type -> Type where
  MkSeq: (1 _: a -> LState s b) -> Sequential s a b
  
runSeq: (1 _: Sequential s a b) -> (a -> LState s b)
runSeq (MkSeq f) = f
\end{idrblk}
where \verb|LState s b| is the state monad with its state bound with linearity:
\begin{idrblk}
data LState: Type -> Type -> Type where
  LST: (1 sf: (1 st: s) -> LC s a) -> LState s a
\end{idrblk}
Hence, the interpreter implemented here maps sequential components of type \verb|seq s a b| to functions of the type \verb|a -> (LState s b)| \footnote{By substituting \Verb|sf| in the definition of \Verb|LState|, this type will be transformed to \Verb|(x: a) -> (1 st: s) -> LC b s|, which is exactly the type we used in the example in \Cref{sec:preliminary:qtt} as the solution.}, which is recognised as the \textit{Kleisli arrow} of the state monad \cite{hughes2005programming}.

By leveraging QTT and interpreting \verb|seq s a b| as the Kleisli arrow of the linear state monad, most of the glue components' implementations are \textit{uniquely}\footnote{By means of how states are used.} determined by their symantics (type signatures).
This is what we mean in \Cref{sec:sym:seq} that \textquote{so that states produced by/consumed from them can be properly handled and, hence, backward path(s) can be properly introduced}.
For instance, the only implementation of \verb|(=<<)| is:
\begin{idrblk}
(=<<) (MkSeq g) (MkSeq f) = 
    MkSeq $ \x => LST $ \st => 
      let LST f'  = f x 
          st' # y = f' st
          LST g'  = g y
      in g' st
\end{idrblk}
or its equivalent variants because the linearity on \verb|st| forces the order of the sequence of let bindings.

\begin{figure}[htbp]
\begin{codeblksmall}
-- interpret dly as a function
dly_fn: {auto sIsState: St s} -> {auto aIsSig: Sig a}
     -> {auto similar: SameShape a s}
     -> (a -> LState s a)
dly_fn = runSeq $ dly reg

-- define the meaning of firing
step: (a -> LState s b)
  -> a -> s -> LC s b
step f x st = runState (f x) st

unrestrict: (!* a) -> a
unrestrict (MkBang unrestricted) = unrestricted

{- if dly has the desired behaviour on each part of the input 
   and corresponding sub-state;
   then dly has the desired behaviour on the composed input and state.-} 
dly_lemma: {auto s1IsState: St s1} -> {auto s2IsState: St s2}
        -> {auto aIsSig: Sig a}    -> {auto bIsSig: Sig b}
        -> {auto similar1: SameShape a s1} 
        -> {auto similar2: SameShape b s2}
        -> (x1: a) -> (x2: b) -> (st1: s1) -> (st2: s2)
        -> (prf1: (step dly_fn x1 st1) 
                = (sigToSt x1 # (unrestrict $ stToSig st1)))
        -> (prf2: (step dly_fn x2 st2) 
                = (sigToSt x2 # (unrestrict $ stToSig st2)))
        -> ((step dly_fn (x1, x2) (the (LPair s1 s2) $ st1 # st2)) 
         = (sigToSt (x1, x2) 
            # (unrestrict $ stToSig (the (LPair s1 s2) $ st1 # st2))))
dly_lemma = --omitted

dly_prop: {auto sIsState: St s} -> {auto aIsSig: Sig a}
       -> {auto similar: SameShape a s}
       -> (x: a) -> (st: s)
       -> (step dly_fn x st) 
        = (sigToSt x # (unrestrict $ stToSig st))
dly_prop {similar = U} {sIsState = LU} x () = Refl
dly_prop {similar = (P prfa prfb)} {sIsState = (LP s1 s2)} 
         {aIsSig = (P pa1 pa2)} (x1, x2) (st1 # st2) 
  = let prf1 = dly_prop x1 st1 
        prf2 = dly_prop x2 st2
    in dly_lemma x1 x2 st1 st2 prf1 prf2
dly_prop {similar = BV} {sIsState = LV} x (MkBang st) = Refl
\end{codeblksmall}
    \caption{The specification and proof of that \protect\Verb|dly| always updates its state by the current input and produces its current state as the current output (\protect\Verb|dly\_prop|).}
    \label{fig:dly-prf}
\end{figure}

Being able to interpret SynQ terms as functions in Idris2 allows us to formally specify and prove components' behaviour in dependent types.
For instance, \Cref{fig:dly-prf} illustrates the proof of the forward delay component \verb|dly| on the property that \textit{firing \Verb|dly| for all valid inputs at all possible states will always cause \Verb|dly| produces its current state as the output and enters the next state that is equal to the current input.}
This property is encoded in dependent types in \Cref{fig:dly-prf} as:
\begin{idrblk}
dly_prop: {auto sIsState: St s}                   -- 1
       -> {auto aIsSig: Sig a}                    -- 2
       -> {auto similar: SameShape a s}           -- 3
       -> (x: a) -> (st: s)                       -- 4
       -> (step dly_fn x st)                      -- 5
        = (sigToSt x # (unrestrict $ stToSig st)) -- 6
\end{idrblk}
in which lines 1 to 4 assert the input and state of \verb|dly| and lines 5 and 6 specify that the firing of \verb|dly| (line 5, where \verb|dly_fn| is the function in Idris2 mapped from \verb|dly|) and its expected result (line 6).

\subsection{The Interpreter to Typed Netlist}
\label{sec:interp:nl}

Netlists (directed graphs) provide a useful view of programs/systems based on what optimisations, e.g. expression sharing by structural equivalence \cite{kiselyov2011implementing}, and code generation can be conducted.
In this section, we consider the interpreter that maps terms in SynQ to \textit{typed} netlists whose input/output ports and compositions are well-typed.
To further illustrate its capability, in this paper, generated netlists are directly translated into synthesisable Verilog HDL code and then visualised by open-source tools Yosys\footnote{https://yosyshq.net/yosys/} and netlistsvg\footnote{https://github.com/nturley/netlistsvg}.

A netlist consists of a list of ordered pairs of labels and instantiated modules (nodes).
Hence, the interpreter for netlists generation must be \textit{context-sensitive} because labels and names of instantiated modules \textit{shall be unique}.
In our case, this context is introduced as a \textit{state} of the interpreter and captured, again, by the state monad.
In practice, different variants of state monads are employed for combinational and sequential components, respectively.

For combinational components, netlists are defined by the type:
\begin{idrblk}
record CombNL a b where
  constructor MkCNL
  iPort: TPort a
  oPort: TPort b
  assignedPorts: List PortAssign
  instModules  : List ModuleInst
\end{idrblk}
which consists of \textit{typed} input and output ports (labels) and \textit{untyped} label pairs and instantiated modules.
Yet the latter two are untyped; they are all generated from typed ports.
In this way, we guarantee that a graph of this type is well-typed.

\begin{figure}[t!]
    \centering
\begin{subfigure}{\linewidth}
\includegraphics[width = \linewidth]{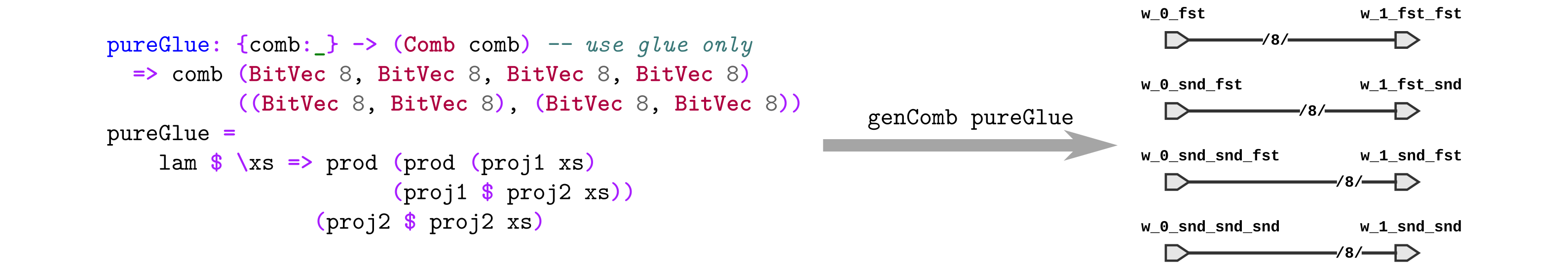}
        \caption{}
        \label{fig:viz-glue}
    \end{subfigure}%
\vskip 1em
    \begin{subfigure}{\linewidth}
    \centering
\includegraphics[width = \linewidth]{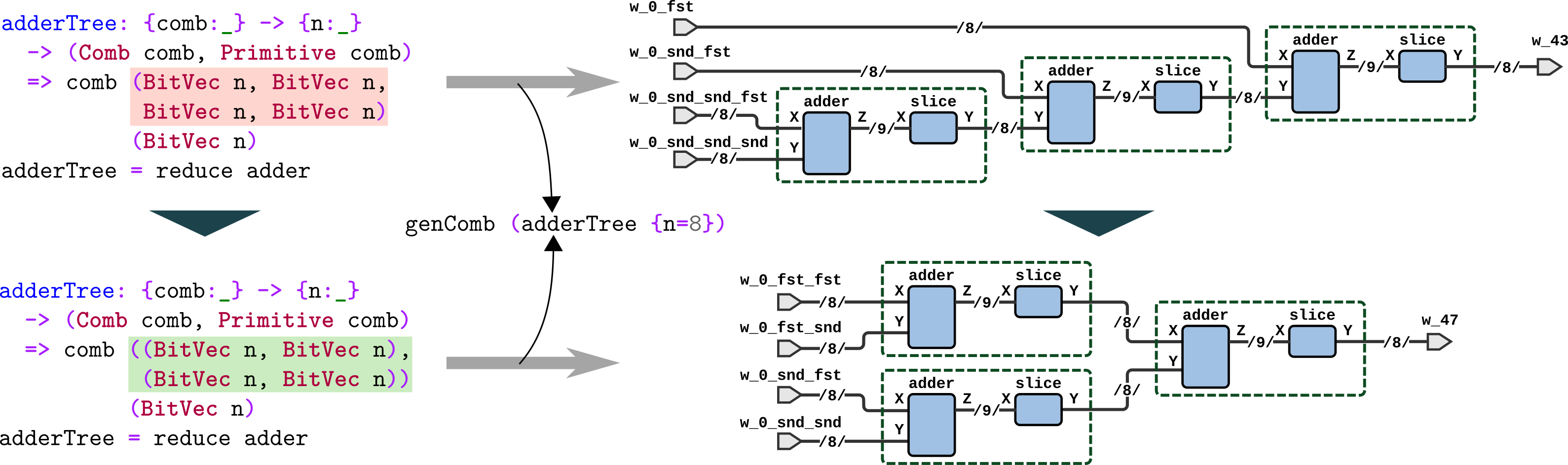}
        \caption{}
        \label{fig:viz-btree}
    \end{subfigure}
    \caption{The visualization of the netlist interpretation of a glue component (a) and \protect\Verb|reduce| in different contexts (b), in which \protect\Verb|adder| is defined as \protect\Verb[commandchars=\\\{\}]|adder = (slice 0 n) << (lam (\textbackslash x => add (proj1 x) (proj2 x))| of type \protect\Verb|comb (BitVec n, BitVec n) (BitVec n)|.}
    \label{fig:viz}
\end{figure}

The interpreter for combinational components is defined by the type constructor:
\begin{idrblk}
record CombinationalNL a b where
  constructor MkComb
  genComb: State Nat (CombNL a b)
\end{idrblk}
where \verb|State| is the conventional (unrestricted) state monad.
By this type constructor, a natural number is employed as the interpreter's state based on what unique labels are generated.
With this interpreter implemented, we can now better illustrate the intuition behind the design of SynQ by visualising interpreted netlists.
For instance, glue components in SynQ are indeed glues defined in \cite{sifakis2015system} because they are simply edges in the netlist view (\Cref{fig:viz-glue}).
Meanwhile, we can now show that our implementation of \verb|reduce| produces different structures in different contexts, as we expected (\Cref{fig:viz-btree}).

Sequential components are interpreted to netlists of the following type:
\begin{idrblk}
data NetList: Type -> Type -> Type -> Type where
  MkNL: (iPort: TPort a) -> (oPort: TPort b)
     -> (1 lPort: LPort s)
     -> (assignedPorts: List PortAssign)
     -> (instModules  : List ModuleInst)
     -> NetList s a b
\end{idrblk}
in which \verb|lPort| bound with multiplicity 1 is used for backward paths of feedback looks.
Consequently, the following type constructor is employed:
\begin{idrblk}
record SequentialNL s a b where
  constructor MkSeq
  1 genSeq: LState2 Nat (NetList s a b)
\end{idrblk}
where \verb|LState2| is:
\begin{idrblk}
data LState2: Type -> Type -> Type where
  LST2: (1 _: s -> LC a s) -> LState2 s a
\end{idrblk}
Compared to \verb|LState| introduced in the previous section, here, the linearity is used to restrict the \textit{entire} state monad instead of its state because we want the generated netlist, especially the feedback loop, to be managed during the interpretation process.

\begin{figure}[t!]
    \centering
\begin{subfigure}{0.8\linewidth}
\includegraphics[width = \linewidth]{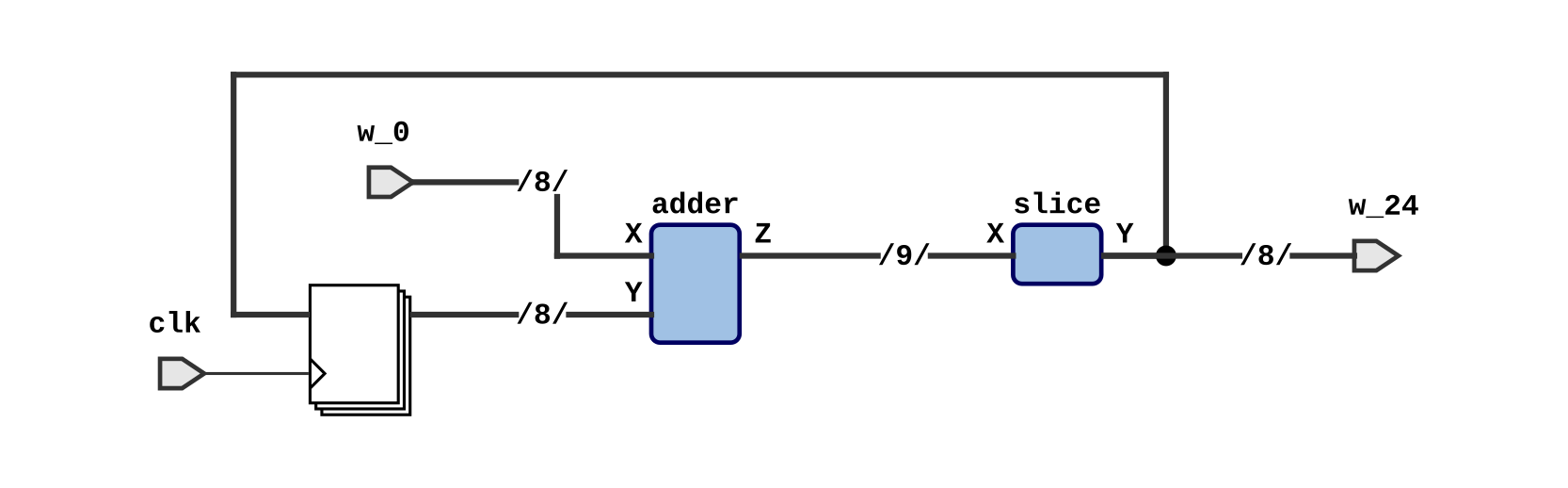}
        \caption{}
        \label{fig:viz-acc-raw}
    \end{subfigure}%
\vskip 1em
    \begin{subfigure}{0.8\linewidth}
    \centering
\includegraphics[width = \linewidth]{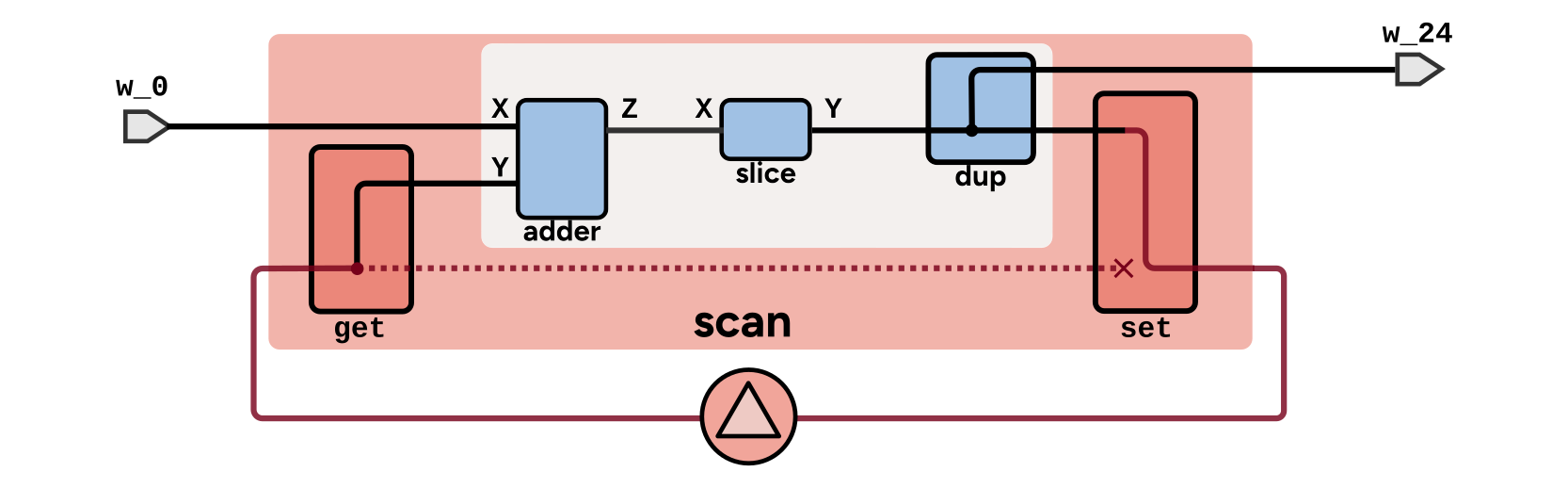}
        \caption{}
        \label{fig:viz-acc-reform}
    \end{subfigure}
    \caption{The visualization of an accumulator's netlist directly generated by open source tools (a) and the netlist's correspondence to the accumulator's specification (b).}
    \label{fig:viz-acc}
\end{figure}

As an example, we generate the netlist of the accumulator specified as follows:
\begin{idrblk}
acc: ... -> seq (!* (BitVec n)) (BitVec n) (BitVec n)
acc = let dup   = lam $ \x => prod x x
          adder = lam $ \x => add (proj1 x) (proj2 x)
      in scan (dup << (slice 0 n) << adder)
\end{idrblk}
The direct visualization of the Verilog HDL code translated from the netlist is illustrated in \Cref{fig:viz-acc-raw}.
By slightly modifying this visualisation and making the implementation of get/set primitives explicit, it can be transformed into the one illustrated in \Cref{fig:viz-acc-reform}, which strictly corresponds to its specification in SynQ.

\subsection{The Interpreter to Symantics}
\label{sec:interp:trans}

Interpreters introduced previously map terms in SynQ to concrete objects.
Consequently, a term that has been interpreted cannot be further interpreted.
In other words, these interpreters make us exit the design flow defined by SynQ.
However, system design processes, as presented in \Cref{fig:design-flow-c} and discussed in \Cref{sec:intro}, often require multiple successive steps so that concerns can be addressed separately.

Here, we illustrate interpreters that interpret terms' symantics by the same or another set of symantics.
With these interpreters, an interpreted term can be further interpreted using the same methodology.
To this end, we can have a coherent multi-step design process based on SynQ and the tagless final approach.

Specifically, the interpreter introduced here performs the following transformation on combinational terms in SynQ:
\begin{equation*}
    \verb|app (lam $ \x => f x) z| \:\: \longmapsto \:\: \verb|f z|
\end{equation*}
that eliminates paired \verb|app| and \verb|lam| in a term when it is applied.
This transformation is referred to as the \textit{normalisation} process, which gives us the ground for asserting equivalence of functional behaviour between terms.
Meanwhile, the existence of a (total) function implementing normalisation is also a witness of SynQ's restricted expressiveness that was discussed in \Cref{sec:sym:comb}.

\begin{figure}
    \centering
    \includegraphics{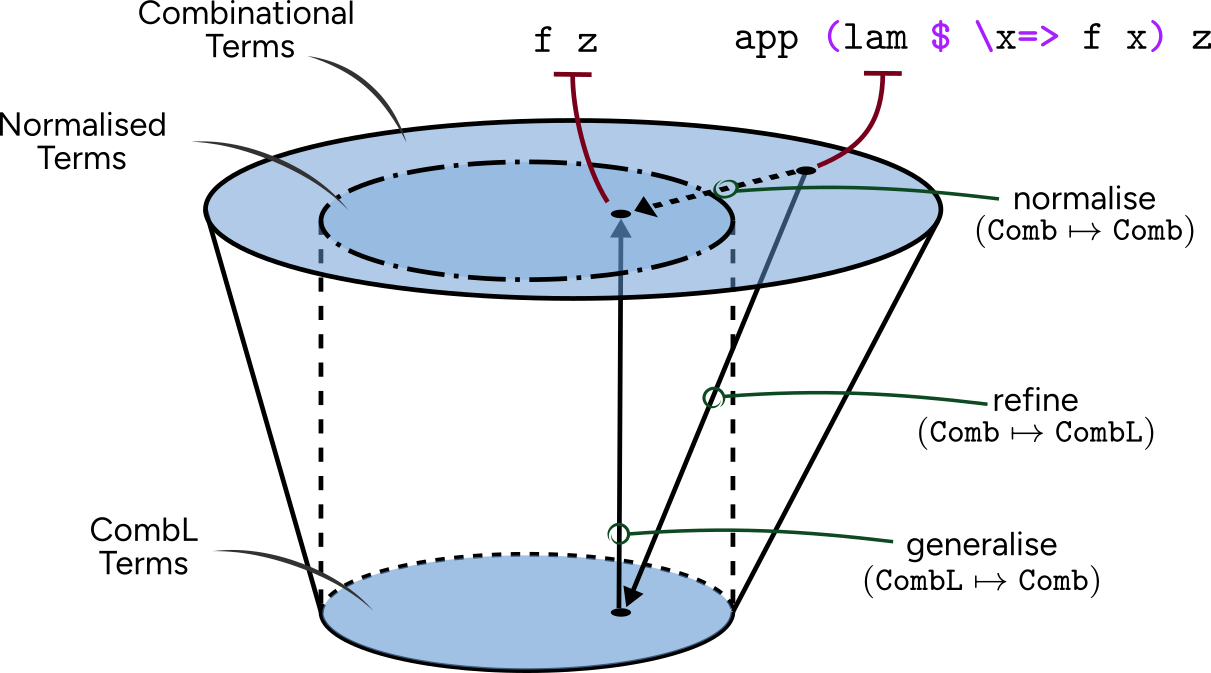}
    \caption{The visualization of interpreters for normalisation, refinement and generalisation discussed in \Cref{sec:interp:trans}, in which solid arrows indicate interpreters that are actually implemented and the dashed arrow indicates the interpreter obtained by composing other interpreters.}
    \label{fig:transform-refine}
\end{figure}

Instead of directly implementing the interpreter for normalisation, here we employ a pair of interpreters for \textit{refinement} and \textit{generalisation} and construct the normalisation interpreter by composing them.
By refinement, terms specified in a language are interpreted by another language that is \textit{less expressive}, i.e., a language with less expressive constructors or with fewer constructors.
For instance, transforming a combinational logic implemented with $\{\land, \lor, \neg\}$ to the one implemented with $\{\text{NAND}\}$ forms a language refinement.
Specifically, we will refine the combinational layer of SynQ to the symantics of another EDSL named \textit{CombL} (combinational less).
Generalisation is the opposite of refinement, which is trivial in our case because the source domain to be generalised is a proper subset of the target domain.
In this way, besides the capability of defining transformations coherently in SynQ, two extra possibilities that
we can: 
(1) coherently introduce other abstraction levels in a SynQ-based design flow; 
and (2) define a SynQ-based design flow by composing atomic design steps in which concerns are properly separated,
are also demonstrated here.
\Cref{fig:transform-refine} sketches our idea of constructing normalisation by composing refinement and generalisation.

The symantics of CombL is similar to the symantics of the combinational layer of SynQ presented in \Cref{fig:synq-symantics-comb}, except that the constructor \verb|app| does not exist in CombL.
Consequently, terms in CombL are all in the normalised form.
On the other hand, since \verb|app| in SynQ enables us to use a \textit{non-atomic} component as a black box, the absence of \verb|app| makes CombL lack the ability of modularisation.
For instance, the accumulator introduced earlier has to be specified as follows if CombL is employed as the language of the combinational layer:
\begin{idrblk}
acc = scan (lam $ \x => prod (slice 0 n $ add (proj1 x) (proj2 x))
                             (slice 0 n $ add (proj1 x) (proj2 x)))
\end{idrblk}
because the definition of the glue component \verb|(<<)| depends on \verb|app|.
This fact suggests that CombL is not a proper top-level language for system design.

Interpreting combinational components by constructors in CombL is conducted by interpreting \verb|app| as the function application in the meta-language Idris2.
In other words, the interpreter \textit{partially} evaluates a term in SynQ in a similar way as the interpreter presented in \Cref{sec:interp:fn} does and directly maps the rest of the term to the corresponding constructor in CombL.
The first step of implementing such an interpreter is \textit{distinguishing} terms that need to be evaluated from terms that will be directly mapped, which is achieved by the following type constructor:
\begin{idrblk}
data E: (Type -> Type -> Type) -> Type -> Type -> Type where
  F: {auto aIsSig: Sig a} -> {auto bIsSig: Sig b}
    -> (E comb () a -> E comb () b) -> E comb a b
  C: comb () a -> E comb () a
\end{idrblk}
This type constructor itself can be considered as a mark in type signatures so that we can distinguish terms in CombL, which is of type \verb|comb a b|, from terms in SynQ, which is typed by \verb|E comb a b|.
It consists of two data constructors.
The first data constructor, \verb|F|, keeps a function in SynQ of type \verb|E comb () a -> E comb () b| as it is so that it can be later applied in the meta-language when an \verb|app| constructor is encountered.
And the second, \verb|C|, states that all values in CombL are also values in SynQ.
With this type constructor defined, the core of implementing the combinational layer of SynQ by CombL is the following:
\begin{idrblk}
{comb:_} -> (CombL comb) => Comb (E comb) where
  lam f = F f
  app (F f) e = f e
  app (C c) e = C c
  ... 
\end{idrblk}
in which \verb|(CombL comb) => Comb (E comb)| states that if \verb|comb| implements CombL, then \verb|E comb| is the implementation of Comb.
The first case of \verb|app| replaces an \verb|app| constructor in SynQ that can be reduced as a function application in meta-language.
Meanwhile, if the first parameter of an \verb|app| is a value that cannot be evaluated (the second case), then the value is simply returned.
Because of that, the second parameter must be the unit in such a case, and \verb|c| can be safely treated as a constant.
Finally, the following function applies the refinement:
\begin{idrblk}
refine: {comb: _} -> (CombL comb) 
     => E comb a b -> comb a b
refine (F f) = lam $ refine . f . C
refine (C x) = x
\end{idrblk}
which sends functions and constants in SynQ to functions and constants in CombL.

The interpreter generalising a CombL term as a SynQ term is just a direct mapping from CombL to Comb.
Hence, we can employ the following type constructor for distinguishing terms in different languages:
\begin{idrblk}
data E0: (Type -> Type -> Type) 
      -> Type -> Type -> Type where
  U: comb a b -> E0 comb a b
\end{idrblk}
which states simply that all terms specified in CombL are terms in the combinational layer of SynQ.
With CombL implemented by \verb|E0 comb| (omitted here) the function that performs the generalisation is given as
\begin{idrblk}
generalise: E0 comb a b -> comb a b
generalise (U x) = x
\end{idrblk}
which simply erases the tag of the data constructor.

\begin{figure}
    \centering
    \includegraphics[width=0.8\linewidth]{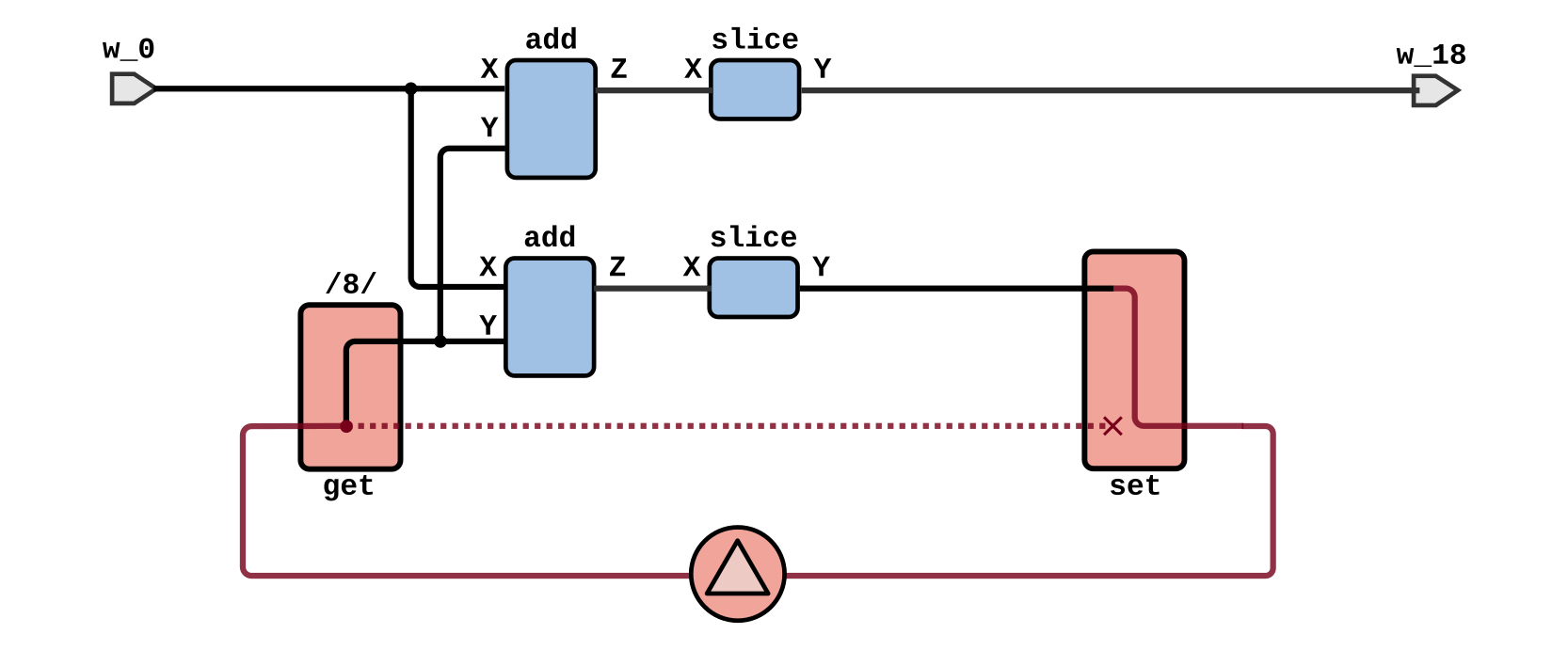}
    \caption{The (reformed) netlist of the accumulator with the normalised combinational part.}
    \label{fig:viz-acc-normed}
\end{figure}

The normalisation interpreter is then defined by the composition of these interpreters:
\begin{idrblk}
normalise: {comb: _} -> (Comb comb) 
        => E (E0 comb) a b -> comb a b
normalise = generalise . refine 
\end{idrblk}
Since \verb|comb| denotes interpreters of the combinational layer of SynQ, \verb|(E0 comb)| and \verb|E (E0 comb)| denote the interpreter of CombL and SynQ, respectively.
Hence, \textit{normalise} forms a transformation within SynQ, as presented in \Cref{fig:transform-refine}.
As a transformation, it can be applied to \textit{any} combinational components in a system specification.
For instance, an accumulator with the normalised combinational part can be specified as:
\begin{idrblk}
acc' = let ...
       in scan (normalise $ dup << (slice 0 n) << adder)
\end{idrblk}
As a specification in its symantics, it can be interpreted to the netlist presented in \Cref{fig:viz-acc-normed} by the interpreter presented in \Cref{sec:interp:nl}, which matches the accumulator specified with CombL presented earlier in this section.

\section{Case Study}
\label{sec:examples}
\begin{figure*}[t]
   \centering
    \includegraphics[width=\linewidth]{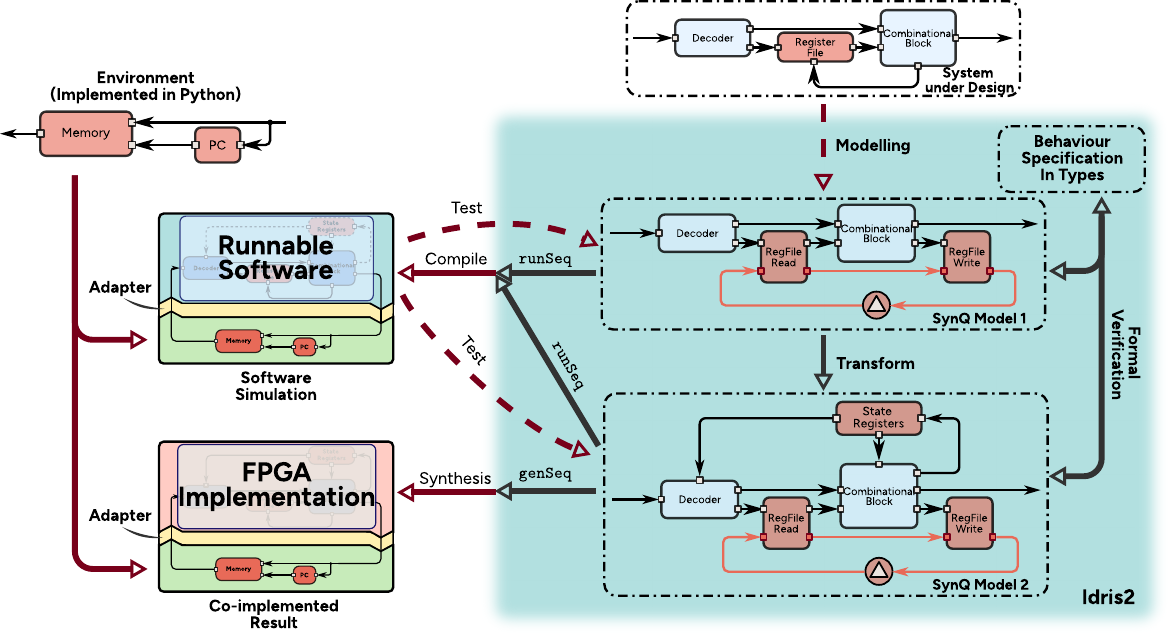}
    \caption{The overview of the case study presented in \Cref{sec:examples}, in which red arrows indicate operations that are outside the SynQ-based design process.}
    \label{fig:case-over-view}
\end{figure*}

We present a case study, which is publicly available in the GitHub repository \cite{omitted}, to further demonstrate how SynQ enables a coherent design process.
In this case study, we consider the specification and implementation of the RV32I instruction set architecture (ISA).
This case study is, in general, aligned with \cite{chen2024qtt}. 
Besides that, enabled by SynQ, most of the design steps that were manually conducted in \cite{chen2024qtt}, e.g. code generation, are now formed as functions in Idris2 and are conducted automatically.
Note that this case study does not target the full implementation of the ISA.
Instead, we assume the existence of memory and program counter (PC), which is considered as the \textit{environment}, and employ a SynQ-based design process to design the rest of the ISA, which forms the \textit{system} reacting to the environment.
In this way, the system to be designed is, in essence, a synchronous reactive system.

To this end, even though an ISA is not typically a software application, the system under design is of the same form as the software applications that SynQ is targeting, which makes this case study sufficiently illustrative.
Meanwhile, the ISA has a well-documented behaviour, which allows us to focus on the formal specification and verification in a type system without discussing how specifications are identified from the requirement.
Further, the ISA also has sufficiently complex functional behaviour.
Hence, by modelling the ISA, the expressiveness of SynQ is implicitly evaluated.

In this case study, a design process, which is mainly hosted by Idris2 as illustrated in \Cref{fig:case-over-view}, is conducted.
This process starts with a naive model of the ISA in SynQ that has a single-stage implementation of \textit{load/store} instructions.
This model is then interpreted as a function in Idris2 glueing black-box components introduced by FFIs, which is compiled and tested under the environment implemented in Python with a manually implemented, lightweight adapter and verified with respect to the dependent-type-based specification(s) of the ISA's functional behaviour.
Since this model has a non-trivial assumption on the environment that \textit{whenever a load/store instruction is produced, so is the data at the corresponding location}, it is then \textit{manually} transformed to a model with staged load/store implementation and the relaxed assumption on the environment.
The transformed model is also tested in the environment and verified \textit{incrementally} in Idris2.
Finally, the verified model is interpreted as Verilog HDL code, synthesised and integrated with the environment on the PYNQ platform\footnote{https://www.pynq.io/}, in which the implementation's behaviour is validated.
In the rest of this section, we will briefly discuss system modelling in SynQ, coherently simulating and implementing the system by leveraging different interpreters and practically conducting formal verification based on the case study. 

\subsection{System Modelling}

The system under design in the case study is modelled in SynQ as follows. 
The initial model, as presented in \Cref{fig:case-over-view}, consists of a \textit{zero-delay} feedback loop from the combinational block to the register file.
Such a non-causal loop is forbidden in SynQ. 
Hence, the first step is to eliminate this feedback loop.
This is accomplished by identifying and decomposing the register file into two components, \verb|RegFile Read| and \verb|RegFile Write|, that can be invoked independently yet share the state.
As shown in \Cref{fig:case-over-view}, this decomposition replaces the zero-delay feedback loop in the initial model with a backward path with a unit delay between two components of the register file, which then gives us the model in SynQ.

Most of the sub-components in both SynQ models before and after the transformation can be given straightforwardly, including state registers in the second model, which can be easily introduced by the \verb|scan| pattern as shown in \Cref{fig:scan}.
However, the register file consists of stateful components, which should be restricted by linearity in the context.
It makes it more challenging to implement.
Here, we show how a \textit{modular}, \textit{reusable}, and \textit{general} register file can be introduced in SynQ with the aid of functions (macros, cf. the remark of \Cref{sec:sym:comb}) in the meta-language Idris2.

To achieve modularisation, we organise components of the register file in a similar way to how components in \verb|Reg| (\Cref{fig:synq-symantics-seq-atomic}) are organised.
That is, a register file is introduced by the interface \verb|RegFile| shown below:
\begin{idrblk}
interface RegFile (comb: Type -> Type -> Type)
                  (seq: Type -> Type -> Type -> Type) where
  constructor MkRF
  1 read: (idx1: comb () (BitVec 5)) -> (idx2: comb () (BitVec 5))
       -> seq RegF () (BitVec 32, BitVec 32)
       
  1 write: (idx: comb () (BitVec 5)) -> (val: comb () (BitVec 32))
        -> seq RegF () ()
\end{idrblk}
This interface indicates that: (1) components consisting in a register file are dependent on both combinational and sequential components; (2) these components have a state (backward path with unit delay) of type \verb|RegF|; and
(3) these components are considered as \textit{resources} as indicated by the multiplicity 1 denoted in front of them.
By introducing the register file in this way, it can be used in a resource-aware manner with always well-formed feedback loops.
Note that the implementation of the \verb|RegFile| interface is based on other components' symantics, i.e., the only implementation of it is of the form:
\begin{idrblk}
{comb:_} -> {seq:_} 
  -> (Primitive comb, Seq comb seq, Reg comb seq) 
  => RegFile comb seq where
  read  = ?read_impl
  write = ?write_impl
\end{idrblk}
with what any interpreter of SynQ will interpret constructors in \verb|RegFile| correctly.

Reusability and generality are facilitated by using generic functions to generate the register file with a specific configuration.
For instance, the \verb|RegF| type is defined as following:
\begin{idrblk}
RegF: Type
RegF = repeatL 32 $ BitVec 32
\end{idrblk}
where \verb|repeatL| is a function defined as:
\begin{idrblk}
repeatL: (n: Nat) -> Type -> Type
repeatL 0 ty = ()
repeatL (S 0) ty = (!* ty)
repeatL (S (S k)) ty = LPair (!* ty) (repeatL (S k) ty)
\end{idrblk}
that produces types satisfying the \verb|St| predicate presented in \Cref{sec:sym:seq}.
With \verb|repeatL|, types of register files with arbitrarily many registers can be introduced.
And the core of indexing an arbitrarily large, non-empty register file can then be given as follows:
\begin{idrblk}
regSel': ... => {n:_} -> {auto prf: LTE 1 n} 
      -> (idx: comb () (BitVec k)) 
      -> (regs: comb () (repeat n $ BitVec 32))
      -> comb () (BitVec 32)
... -- case one register only
regSel' {n = (S (S k))} {prf = (LTESucc prf')} idx regs = 
  let cur = const $ BV $ cast (S k)
      (cur_val, rest) = unpack {n = S k} regs
  in mux21 (eq idx cur) cur_val (regSel' idx rest)
\end{idrblk}
Here the component is \textit{inductively} constructed by pattern matching on \verb|n| and \verb|repeat| is another type level function generating types for the \verb|Sig| predicate (\Cref{sec:sym:comb}).

\subsection{Multiple Interpretations of The System}

The ``symantics-first'' modelling enabled by SynQ and the glue/atomic-component partitioning of SynQ allows us to make platform-independent models, which can be implemented on different platforms by employing different interpreters and/or changing the underlying implementations of components.
Specifically, in this case study, the software implementation, which is used for simulating and testing the functional behaviour of the system, is obtained by interpreting glue components as Idris2 functions that glue atomic components implemented in C and introduced by FFIs.
Meanwhile, the hardware (FPGA) implementation is obtained by a direct and automatic translation of the resulting typed netlist generated from the interpreter introduced in \Cref{sec:interp:nl}.
This implementation relies on \textit{only} the interface of atomic components, which allows us to further decouple a component's specification and its implementation.
In this way, how SynQ can be used together with legacy code or black-box components and for code generation are respectively demonstrated.

As illustrated in \cite{chen2024qtt}, one benefit of adopting a component-based design framework is that the same component with different underlying implementations can be used in the same environment.
As presented by \Cref{fig:case-over-view}, using different implementations of a model (the transformed model in this case study) in the same environment can be simply conducted by invoking the corresponding interpreters, sending artefacts generated from interpreters to successive platform-dependent steps (compiling/synthesising in the case study) and putting them into the environment with proper adapters (cf. \cite{omitted} for implementation details).
To this end, we are allowed to obtain the consistency between the behaviour of different implementations of the same model in SynQ.
Specifically, in this case study, this consistency is leveraged to perform an efficient simulation of the model so that the correctness of the final implementation can be assured.

\subsection{Formal Verification}

Formal verification of the SynQ model in this case study follows the same methodology as presented in \Cref{fig:dly-prf}.
However, special attention needs to be paid to two practical issues.
Firstly, the speed of the type-checker prevents us from interactively conducting an end-to-end proof of the entire system's functional behaviour.
Hence, the verification in the case study focuses only on the combinational part of the system. 
The rest of the system, e.g., the register file or state registers, is used under the restriction of linearity, which allows us to reason their behaviour easily by their algebraic specification.

Secondly, adopting the component-based design framework prevents us from fully verifying the models' behaviour in the type system.
This is because an atomic component in a component-based design framework is specified by its interfaces (as a black box), whereas reasoning about the behaviour of such a component in a type system often requires details about its implementation.
In this case study, the formal verification is conducted in two steps.
Firstly, in Idris2, the property to be proved is formally specified in types and proved on the interpreted SynQ model by identifying and introducing \textit{assumptions} on the behaviour of atomic components.
Secondly, identified assumptions are used as \textit{guarantees} and verified on the implementation of atomic components being used in the domain where the implementation is made.

For example, we demonstrate here how a part of the \textit{add} instruction is verified.
The property to be verified here is that \textit{if an add instruction parametrised with indices \(\mathtt{rs1}\) and \(\mathtt{rs2}\) of the register file is given, then the data to be written to the register file equals to the lower 32 bits of the sum of the data at the position \(\mathtt{rs1}\) and \(\mathtt{rs2}\) in the register file.}
The top-level specification is then given as:
\begin{idrblk}
prop_add_rdata: (ins: InTy) 
  -> getOutRData (combAddFn ins) 
     = (bvSlice 0 32 $ bvAdd (fst $ getInRData ins) 
                             (snd $ getInRData ins))
\end{idrblk}
in which \verb|combAddFn| is the interpreted SynQ model \textit{partially specialisted} with known fields (opcode, funct3, funct7, etc.) in the add instruction and \verb|ins| is the rest of the instruction (rd, rs1, rs2) on what the for all quantifier is applied.
An intermediate step of the proof is that:
\begin{idrblk}
prop_add_rdata ins = rewrite assume_or_1 {n=1} in 
                     rewrite assume_slice in ?rhs_rdata
\end{idrblk}
in which some assumptions have been identified. 
The target type on the hole \verb|?rhs_rdata| is then:
\begin{idrblk}
         ins : ...
----------------------------------------------------------------
 "rhs_rdata" : (if bvOr (BV 1) (BV 0) == BV 1 then...else...)... 
                 = bvSlice 0 32 (bvAdd (...) (...))
\end{idrblk}
which indicates that the LHS of the target type cannot be further reduced because the type checker cannot infer the value of the condition of the if-statement that \verb|bvOr (BV 1) (BV 0) == BV 1|.
We can then introduce a reasonable assumption as follows:
\begin{idrblk}
assume_or_2: {n:_} -> {auto _: LTE 1 n} 
  -> (bvOr {n=n} (BV 1) (BV 0)) = BV {n=n} 1
\end{idrblk}
which states that for all bit vectors whose length is not zero, the result of \verb|bvOr (BV 1) (BV 0)| is always equal to \verb|BV 1|.
With this assumption, the proof is finished as follows:
\begin{idrblk}
prop_add_rdata ins = rewrite assume_or_2 {n=1} in
                     rewrite assume_or_1 {n=1} in 
                     rewrite assume_slice in Refl
\end{idrblk}
When the C implementation of \verb|bvOr| is used, this assumption would give us the following contract in ACSL:
\begin{idrblk}
/*@ requires len > 0;
  @ assigns \nothing;
  @ ensures (val_1 == 1) && (val_2 == 0) ==> (\result == 1);
  @ */
uint64_t bv_or(uint8_t len, uint64_t val_1, uint64_t val_2);
\end{idrblk}
in which \verb|ensures| is exactly the assumption introduced by \verb|assume_or_2|.
Note that, in this case, we are translating specifications in Idris2 to specifications in ACSL, which are both in the same abstraction level.
Hence, the semantics gap from specifications to systems' behaviour models is minimised.

\section{Related Work}
\label{sec:related}
Many of the related works have been discussed in early sections. 
Here, we review this research in a more organised manner and also introduce related works that have not been discussed yet.
From the \textit{system design methodology} perspective, a representative instance of component-based design is the BIP (\textbf{B}ehaviour, \textbf{I}nteraction, \textbf{P}riority) framework \cite{gossler2005composition,basu2008bip}.
In the BIP framework, atomic components are labelled transition systems, and systems are constructed by restricting the product of components.
This difference in how components are modelled and glued makes the design methodology supported by SynQ and BIP two distinct methodologies following the same component-based design principle according to \cite{sifakis2015system}.
The design methodology entailed by the EDSL is also related to platform-based and contract-based design \cite{keutzer2000system,sangiovanni2007quo,sangiovanni2012taming,benveniste2018contracts}.
Even though conducting platform (language) refinements in SynQ has been briefly demonstrated (\Cref{sec:interp:trans}), more investigations are desired to formally introduce these concepts.

From the \textit{language perspective}, SynQ and other synchronous programming languages, such as Lustre \cite{pilaud1987lustre}, Signal \cite{benveniste1991signal} and Esterel \cite{wester2013space}, are targeting the same set of systems.
The largest distinction between SynQ and these languages is that SynQ is designed in the ``symantics-first'' style, which makes it closer to a proof-of-concept on how synchronous systems in QTT are characterised instead of an implementation of a semantics of synchronous systems.
A comprehensive review of the early stages of these languages can be found in \cite{benveniste2003synchronous}.
Later, the Vélus project \cite{bourke2017formally,bourke2023verified}, combined with CompCert \cite{leroy2009Compcert}, achieves a fully verified compilation flow from Lustre to runnable programs.
Meanwhile, the MARVeLus (\textbf{M}ethod for \textbf{A}utomated \textbf{R}eﬁnement-type \textbf{Ve}riﬁcation of \textbf{Lus}tre) project \cite{chen2022synchronous,chen2024synchronous} shows that properties of Lustre program can be specified in refinement types and formally verified.
Combining MARVeLus, Vélus, and CompCert will then, to some extent, give us the desired coverage of the full three stages of a design process.
However, the expressiveness of refinement types is relatively more restricted than QTT, and platforms supported by CompCert are also limited.
Our investigation on SynQ is indeed more ambitious and targets systematic methodologies for embedded system design.

Finally, if we restrict the target implementations of systems to synchronous digital circuits, how functional programming languages can be used has been investigated since the early 1980s, which was surveyed in \cite{gammie2013synchronous}.
Note that there is also research which is relevant to the topic but out of the scope of \cite{gammie2013synchronous}, such as C$\lambda$ash \cite{baaij2010c}, which reuses Haskell's frontend, and Chisel \cite{bachrach2012chisel}, which embeds its frontend into Scala.
Among these works, \cite{megacz2011hardware} addresses challenges, specifically \textit{disallowing higher-order functions in circuits} and \textit{distinguishing recursive structure from feedback}, which are also part of concerns addressed by SynQ.
These challenges are addressed in \cite{megacz2011hardware} by $\kappa$-calculus \cite{hasegawa1995decomposing} and in SynQ by type-based contracts and macros (\Cref{sec:sym:comb}).

\section{Conclusion and Discussion}
\label{sec:conclusion}
The contribution of this paper is twofold.
From the \textit{practical} perspective, this paper introduces SynQ, an EDSL embedded in Idris2 via the tagless final approach.
SynQ allows us to design synchronous systems in a component-based framework (\Cref{sec:sym:cbd}) and enables fine-grained behaviour and resource usage specification and control by leveraging the quantitative type theory (\Cref{sec:sym:seq}).
Three interpreters of models in SynQ are introduced, which respectively illustrate how terms in SynQ can be interpreted to: (1) Idris2 functions with black box atomic components (\Cref{sec:interp:fn}), which can be formally verified and directly compiled to runnable software as implementation or for simulation; (2) typed netlists which can be used for optimisation and code generation (\Cref{sec:interp:nl}); and (3) transformed terms in SynQ or refined terms in other EDSLs (\Cref{sec:interp:trans}).
The composability of these interpreters, as illustrated in \Cref{sec:interp:trans}, indicates that a coherent semantics is shared by these interpreters, which is a key requirement for system design automation as sketched by Sifakis \cite{sifakis2015system}.
The case study (\Cref{sec:examples}) evidently presented how a semantic coherent design process can be built based on SynQ, in which formal specification and verification, system modelling, design transformation and implementation can be conducted coherently and seamlessly.
These facts make SynQ a step towards system design automation according to principles proposed by Sifakis \cite{sifakis2015system}.

From the \textit{methodology} perspective, SynQ as a DSL is designed by formally \textit{specialising} its meta-language with type-encoded contracts (\Cref{sec:sym}) and implemented by the (tagless final) \textit{language embedding}, which is less common under the taxonomy of DSL's design and implementation patterns proposed by Mernik et al. \cite{mernik2005DSL}.
On the other hand, SynQ's design and implementation make SynQ a formal system that is formally embedded in the quantitative type theory. 
It is suggested by the logic framework view of the tagless final approach that is briefly sketched in \Cref{sec:preliminary:tf}  and how the \textit{precise} interpretation of a sequential glue component in SynQ is uniquely determined by its symantics/type signature (\Cref{sec:sym:seq,sec:interp:fn} and also the example in \Cref{sec:preliminary:qtt}).
The design and implementation of SynQ presented in early sections then serve as an initial step and a proof-of-concept for a general approach of how design processes can be formally modelled and studied in a type theory.
To this end, by mapping formally embedded design processes back via Curry-Howard correspondence, we can investigate better languages and practical solutions for system design automation.

The future work on both perspectives is intertwined.
Firstly, the implementation of SynQ is still in the proof-of-concept stage, which means a large number of implementations remain.
To fully support a practical design process, more EDSLs similar to \verb|CompL| in \Cref{sec:interp:trans}, which model different abstraction levels where SynQ can be transformed from/to, and interpreters applying transformations/optimisations need to be introduced.
Implementing these ``basic blocks'' of a design process will enable us to obtain automated design processes customised for different target architectures.
In the meantime, it will also allow us to further formalise and analysis design processes.
Secondly, the correspondence between interpreters of SynQ demands formal proof.
In this direction, interpreters to Idris2 functions \Cref{sec:interp:fn} and to typed netlists \Cref{sec:interp:nl} play special roles.
The former, as discussed in \Cref{sec:preliminary:tf}, \textit{denotes} terms in SynQ by Idris2 functions, while the latter gives us interpretations which may be considered in the \textit{algebraicly free} form.
Hence, the proof of their correspondence will draw the boundary of interpreters on the functional behaviour aspect.
Finally, in which type theory (meta-language) design processes should be embedded also deserves further investigation.
The current layered design of SynQ (\Cref{sec:sym:sym}), in essence, uses unrestricted and linear types only.
It makes the type system of SynQ itself closer to a type system with a separated context \cite{krishnaswami2015integrating,vakar2015categorical}.
How the multiplicity 0 could be used in a design process or whether system design can benefit from QTT with the natural number semiring (\cite{atkey2024polynomial}) are open questions that need to be answered in further practice.

%
%
%
\bibliographystyle{splncs04}
\bibliography{ref.bib}

\begin{thebibliography}{10}
\providecommand{\url}[1]{\texttt{#1}}
\providecommand{\urlprefix}{URL }
\providecommand{\doi}[1]{https://doi.org/#1}

\bibitem{omitted}
The repository of {SynQ} \url{https://github.com/ruich95/SynQ.git}

\bibitem{abramsky1993computational}
Abramsky, S.: Computational interpretations of linear logic. Theoretical Computer Science  \textbf{111}(1-2),  3--57 (1993)

\bibitem{atkey2018syntax}
Atkey, R.: Syntax and semantics of quantitative type theory. In: Proceedings of the 33rd Annual ACM/IEEE Symposium on Logic in Computer Science. pp. 56--65 (2018)

\bibitem{atkey2024polynomial}
Atkey, R.: Polynomial time and dependent types. Proceedings of the ACM on Programming Languages  \textbf{8}(POPL),  2288--2317 (2024)

\bibitem{baaij2010c}
Baaij, C., Kooijman, M., Kuper, J., Boeijink, A., Gerards, M.: C$\lambda$ash: Structural descriptions of synchronous hardware using {H}askell. In: 2010 13th Euromicro Conference on Digital System Design: Architectures, Methods and Tools. pp. 714--721. IEEE (2010)

\bibitem{bachrach2012chisel}
Bachrach, J., Vo, H., Richards, B., Lee, Y., Waterman, A., Avi{\v{z}}ienis, R., Wawrzynek, J., Asanovi{\'c}, K.: {Chisel: constructing hardware in a Scala embedded language}. In: Proceedings of the 49th Annual Design Automation Conference. pp. 1216--1225 (2012)

\bibitem{basu2008bip}
Basu, A.: {Component-based Modeling of Heterogeneous Real-time Systems in BIP}. Ph.d. theses, {Universit{\'e} Joseph-Fourier - Grenoble I} (Dec 2008)

\bibitem{basu2007using}
Basu, A., Mounier, L., Poulhies, M., Pulou, J., Sifakis, J.: {Using BIP for modeling and verification of networked systems--a case study on tinyos-based networks}. In: Sixth IEEE International Symposium on Network Computing and Applications (NCA 2007). pp. 257--260. IEEE (2007)

\bibitem{baudin2021acsl}
Baudin, P., Filli{\^a}tre, J.C., March{\'e}, C., Monate, B., Moy, Y., Prevosto, V.: {ACSL: ANSI/ISO C specification}. URL https://frama-c. com/html/acsl. html  (2021)

\bibitem{benveniste2018contracts}
Benveniste, A., Caillaud, B., Nickovic, D., Passerone, R., Raclet, J.B., Reinkemeier, P., Sangiovanni-Vincentelli, A., Damm, W., Henzinger, T.A., Larsen, K.G., et~al.: Contracts for system design. Foundations and Trends{\textregistered} in Electronic Design Automation  \textbf{12}(2-3),  124--400 (2018)

\bibitem{benveniste2003synchronous}
Benveniste, A., Caspi, P., Edwards, S.A., Halbwachs, N., Le~Guernic, P., De~Simone, R.: The synchronous languages 12 years later. Proceedings of the IEEE  \textbf{91}(1),  64--83 (2003)

\bibitem{benveniste1991signal}
Benveniste, A., Le~Guernic, P., Jacquemot, C.: Synchronous programming with events and relations: the {SIGNAL} language and its semantics. Science of Computer Programming  \textbf{16}(2),  103--149 (1991)

\bibitem{biernacki2008obc}
Biernacki, D., Cola{\c{c}}o, J.L., Hamon, G., Pouzet, M.: Clock-directed modular code generation for synchronous data-flow languages. In: Proceedings of the 2008 ACM SIGPLAN-SIGBED conference on Languages, compilers, and tools for embedded systems. pp. 121--130 (2008)

\bibitem{boulton1992experience}
Boulton, R.J., Gordon, A.D., Gordon, M.J., Harrison, J., Herbert, J., Van~Tassel, J.: Experience with embedding hardware description languages in hol. In: TPCD. vol.~10, pp. 129--156 (1992)

\bibitem{bourke2017formally}
Bourke, T., Brun, L., Dagand, P.{\'E}., Leroy, X., Pouzet, M., Rieg, L.: A formally verified compiler for lustre. In: Proceedings of the 38th ACM SIGPLAN Conference on Programming Language Design and Implementation. pp. 586--601 (2017)

\bibitem{bourke2023verified}
Bourke, T., Pesin, B., Pouzet, M.: Verified compilation of synchronous dataflow with state machines. ACM Transactions on Embedded Computing Systems  \textbf{22}(5s),  1--26 (2023)

\bibitem{boussinot1991esterel}
Boussinot, F., De~Simone, R.: The esterel language. Proceedings of the IEEE  \textbf{79}(9),  1293--1304 (1991)

\bibitem{bozga2009modeling}
Bozga, M.D., Sfyrla, V., Sifakis, J.: {Modeling synchronous systems in BIP}. In: Proceedings of the seventh ACM international conference on Embedded software. pp. 77--86 (2009)

\bibitem{brady2021idris}
Brady, E.: Idris 2: Quantitative type theory in practice. In: 35th European Conference on Object-Oriented Programming (ECOOP 2021). Schloss Dagstuhl-Leibniz-Zentrum f{\"u}r Informatik (2021)

\bibitem{carette2009finally}
Carette, J., Kiselyov, O., Shan, C.C.: Finally tagless, partially evaluated: Tagless staged interpreters for simpler typed languages. Journal of Functional Programming  \textbf{19}(5),  509--543 (2009)

\bibitem{chen2022synchronous}
Chen, J., Vargas~de Mendon{\c{c}}a, J.L., Jalili, S., Ayele, B., Bekele, B.N., Qu, Z., Sharma, P., Shiferaw, T., Zhang, Y., Jeannin, J.B.: Synchronous programming and refinement types in robotics: From verification to implementation. In: Proceedings of the 8th ACM SIGPLAN International Workshop on Formal Techniques for Safety-Critical Systems. pp. 68--79 (2022)

\bibitem{chen2024synchronous}
Chen, J., de~Mendon{\c{c}}a, J.L.V., Ayele, B.S., Bekele, B.N., Jalili, S., Sharma, P., Wohlfeil, N., Zhang, Y., Jeannin, J.B.: Synchronous programming with refinement types. Proceedings of the ACM on Programming Languages  \textbf{8}(ICFP),  938--972 (2024)

\bibitem{chen2024qtt}
Chen, R., Sander, I.: A quantitative type approach to formal component-based system design. In: 2024 Forum on Specification \& Design Languages (FDL). pp. 1--10. IEEE (2024)

\bibitem{christiansen2016elaborator}
Christiansen, D., Brady, E.: {Elaborator reflection: extending Idris in Idris}. In: Proceedings of the 21st ACM SIGPLAN International Conference on Functional Programming. pp. 284--297 (2016)

\bibitem{coq1996coq}
Coq, P.: The coq proof assistant-reference manual. INRIA Rocquencourt and ENS Lyon, version  \textbf{5} (1996)

\bibitem{gammie2013synchronous}
Gammie, P.: Synchronous digital circuits as functional programs. ACM Computing Surveys (CSUR)  \textbf{46}(2),  1--27 (2013)

\bibitem{ghica2014bounded}
Ghica, D.R., Smith, A.I.: Bounded linear types in a resource semiring. In: Programming Languages and Systems: 23rd European Symposium on Programming, ESOP 2014, Held as Part of the European Joint Conferences on Theory and Practice of Software, ETAPS 2014, Grenoble, France, April 5-13, 2014, Proceedings 23. pp. 331--350. Springer (2014)

\bibitem{gibbons2014folding}
Gibbons, J., Wu, N.: Folding domain-specific languages: deep and shallow embeddings (functional pearl). In: Proceedings of the 19th ACM SIGPLAN international conference on Functional programming. pp. 339--347 (2014)

\bibitem{gilles1974semantics}
Gilles, K.: The semantics of a simple language for parallel programming. Information processing  \textbf{74}(471-475),  15--28 (1974)

\bibitem{girard1987linear}
Girard, J.Y.: Linear logic. Theoretical computer science  \textbf{50}(1),  1--101 (1987)

\bibitem{gordon1993hol}
Gordon, M.J., Melham, T.F.: Introduction to HOL: A theorem proving environment for higher order logic. Cambridge University Press (1993)

\bibitem{gossler2005composition}
G{\"o}ssler, G., Sifakis, J.: Composition for component-based modeling. Science of Computer Programming  \textbf{55}(1-3),  161--183 (2005)

\bibitem{harper1993framework}
Harper, R., Honsell, F., Plotkin, G.: A framework for defining logics. Journal of the ACM (JACM)  \textbf{40}(1),  143--184 (1993)

\bibitem{hasegawa1995decomposing}
Hasegawa, M.: Decomposing typed lambda calculus into a couple of categorical programming languages. In: International Conference on Category Theory and Computer Science. pp. 200--219. Springer (1995)

\bibitem{hoare1969axiomatic}
Hoare, C.A.R.: An axiomatic basis for computer programming. Communications of the ACM  \textbf{12}(10),  576--580 (1969)

\bibitem{hudak2007history}
Hudak, P., Hughes, J., Peyton~Jones, S., Wadler, P.: A history of haskell: being lazy with class. In: Proceedings of the third ACM SIGPLAN conference on History of programming languages. pp. 12--1 (2007)

\bibitem{hughes2005programming}
Hughes, J.: Programming with arrows. In: Advanced Functional Programming: 5th International School, AFP 2004, Tartu, Estonia, August 14--21, 2004, Revised Lectures. pp. 73--129. Springer (2005)

\bibitem{hutton1999fold}
Hutton, G.: A tutorial on the universality and expressiveness of fold. Journal of Functional Programming  \textbf{9}(4),  355--372 (1999)

\bibitem{jones1993system}
Jones, M.P.: A system of constructor classes: overloading and implicit higher-order polymorphism. In: Proceedings of the conference on Functional programming languages and computer architecture. pp. 52--61 (1993)

\bibitem{jones1997type}
Jones, S.P., Jones, M., Meijer, E.: Type classes: an exploration of the design space. In: Haskell workshop. pp. 1--16 (1997)

\bibitem{keutzer2000system}
Keutzer, K., Newton, A.R., Rabaey, J.M., Sangiovanni-Vincentelli, A.: System-level design: Orthogonalization of concerns and platform-based design. IEEE transactions on computer-aided design of integrated circuits and systems  \textbf{19}(12),  1523--1543 (2000)

\bibitem{kiselyov2011implementing}
Kiselyov, O.: Implementing explicit and finding implicit sharing in embedded {DSL}s. In: Proceedings of the IFIP Working Conference on Domain-Specific Languages (DSL'11). pp. 210--225 (2011)

\bibitem{krishnaswami2015integrating}
Krishnaswami, N.R., Pradic, P., Benton, N.: Integrating linear and dependent types. ACM SIGPLAN Notices  \textbf{50}(1),  17--30 (2015)

\bibitem{lattner2004llvm}
Lattner, C., Adve, V.: {LLVM}: A compilation framework for lifelong program analysis \& transformation. In: International symposium on code generation and optimization, 2004. CGO 2004. pp. 75--86. IEEE (2004)

\bibitem{lattner2021mlir}
Lattner, C., Amini, M., Bondhugula, U., Cohen, A., Davis, A., Pienaar, J., Riddle, R., Shpeisman, T., Vasilache, N., Zinenko, O.: {MLIR}: Scaling compiler infrastructure for domain specific computation. In: 2021 IEEE/ACM International Symposium on Code Generation and Optimization (CGO). pp. 2--14. IEEE (2021)

\bibitem{leroy2009Compcert}
Leroy, X.: Formal verification of a realistic compiler. Communications of the ACM  \textbf{52}(7),  107--115 (2009), \url{http://xavierleroy.org/publi/compcert-CACM.pdf}

\bibitem{lindholm2013jvm}
Lindholm, T., Yellin, F., Bracha, G., Buckley, A.: The Java virtual machine specification. Addison-wesley (2013)

\bibitem{mcbride2016got}
McBride, C.: I got plenty o’nuttin’. A List of Successes That Can Change the World: Essays Dedicated to Philip Wadler on the Occasion of His 60th Birthday pp. 207--233 (2016)

\bibitem{megacz2011hardware}
Megacz, A.: Hardware design with generalized arrows. In: International Symposium on Implementation and Application of Functional Languages. pp. 164--180. Springer (2011)

\bibitem{meijer1991functional}
Meijer, E., Fokkinga, M., Paterson, R.: Functional programming with bananas, lenses, envelopes and barbed wire. In: Conference on functional programming languages and computer architecture. pp. 124--144. Springer (1991)

\bibitem{mernik2005DSL}
Mernik, M., Heering, J., Sloane, A.M.: When and how to develop domain-specific languages. ACM computing surveys (CSUR)  \textbf{37}(4),  316--344 (2005)

\bibitem{de2023wiring}
de~Muijnck-Hughes, J., Vanderbauwhede, W.: Wiring circuits is easy as $\{$0, 1, w$\}$, or is it... In: 37th European Conference on Object-Oriented Programming, ECOOP 2023. p.~8 (2023)

\bibitem{nederpelt2014type}
Nederpelt, R., Geuvers, H.: Type theory and formal proof: an introduction. Cambridge University Press (2014)

\bibitem{nielson1992two}
Nielson, F., Nielson, H.R.: Two-level functional languages. Cambridge university press (1992)

\bibitem{petricek2014coeffects}
Petricek, T., Orchard, D., Mycroft, A.: Coeffects: a calculus of context-dependent computation. ACM SIGPLAN Notices  \textbf{49}(9),  123--135 (2014)

\bibitem{pfenning1988higher}
Pfenning, F., Elliott, C.: Higher-order abstract syntax. ACM SIGPLAN Notices  \textbf{23}(7),  199--208 (1988)

\bibitem{pilaud1987lustre}
Pilaud, D., Halbwachs, N., Plaice, J.: Lustre: A declarative language for programming synchronous systems. In: Proceedings of the 14th Annual ACM Symposium on Principles of Programming Languages (14th POPL 1987). ACM, New York, NY. vol.~178, p.~188. Citeseer (1987)

\bibitem{plotkin1981structural}
Plotkin, G.D.: A structural approach to operational semantics  (1981)

\bibitem{polakow2015embedding}
Polakow, J.: Embedding a full linear lambda calculus in haskell. ACM SIGPLAN Notices  \textbf{50}(12),  177--188 (2015)

\bibitem{sangiovanni2007quo}
Sangiovanni-Vincentelli, A.: {Quo vadis, SLD? Reasoning about the trends and challenges of system level design}. Proceedings of the IEEE  \textbf{95}(3),  467--506 (2007)

\bibitem{sangiovanni2012taming}
Sangiovanni-Vincentelli, A., Damm, W., Passerone, R.: Taming {D}r. {F}rankenstein: {C}ontract-based design for cyber-physical systems. European Journal of Control  \textbf{18}(3),  217--238 (2012)

\bibitem{sangiovanni2001platform}
Sangiovanni-Vincentelli, A., Martin, G.: Platform-based design and software design methodology for embedded systems. IEEE Design \& Test of computers  \textbf{18}(6),  23--33 (2001)

\bibitem{schafer2016axiomatic}
Sch{\"a}fer, S., Schneider, S., Smolka, G.: Axiomatic semantics for compiler verification. In: Proceedings of the 5th ACM SIGPLAN Conference on Certified Programs and Proofs. pp. 188--196 (2016)

\bibitem{sfyrla2010compositional}
Sfyrla, V., Tsiligiannis, G., Safaka, I., Bozga, M., Sifakis, J.: {Compositional translation of Simulink models into synchronous BIP}. In: International Symposium on Industrial Embedded System (SIES). pp. 217--220. IEEE (2010)

\bibitem{sifakis2014toward}
Sifakis, J.: Toward a system design science. In: From Programs to Systems. The Systems perspective in Computing: ETAPS Workshop, FPS 2014, in Honor of Joseph Sifakis, Grenoble, France, April 6, 2014. Proceedings. pp. 225--234. Springer (2014)

\bibitem{sifakis2015system}
Sifakis, J.: System design automation: Challenges and limitations. Proceedings of the IEEE  \textbf{103}(11),  2093--2103 (2015)

\bibitem{taha1997multi}
Taha, W., Sheard, T.: Multi-stage programming with explicit annotations. In: Proceedings of the 1997 ACM SIGPLAN symposium on Partial evaluation and semantics-based program manipulation. pp. 203--217 (1997)

\bibitem{terui2001light}
Terui, K.: Light affine lambda calculus and polytime strong normalization. In: Proceedings 16th Annual IEEE Symposium on Logic in Computer Science. pp. 209--220. IEEE (2001)

\bibitem{tov2011affine-types}
Tov, J.A., Pucella, R.: Practical affine types. ACM SIGPLAN Notices  \textbf{46}(1),  447--458 (2011)

\bibitem{turner2004total}
Turner, D.A.: Total functional programming. J. Univers. Comput. Sci.  \textbf{10}(7),  751--768 (2004)

\bibitem{vakar2015categorical}
V{\'a}k{\'a}r, M.: A categorical semantics for linear logical frameworks. In: Foundations of Software Science and Computation Structures: 18th International Conference, FOSSACS 2015, Held as Part of the European Joint Conferences on Theory and Practice of Software, ETAPS 2015, London, UK, April 11-18, 2015, Proceedings 18. pp. 102--116. Springer (2015)

\bibitem{wester2013space}
Wester, R., Kuper, J.: A space/time tradeoff methodology using higher-order functions. In: 2013 23rd International Conference on Field programmable Logic and Applications. pp.~1--2. IEEE (2013)

\end{thebibliography}
%




\end{document}